%% file: main.tex
\documentclass[12pt,a4paper]{article}

\usepackage{amsmath,amsthm,amssymb,amscd,a4wide}
\usepackage[dvips]{graphicx}

\usepackage{dsfont}
\usepackage{mathrsfs}

\usepackage[cmtip,arrow]{xy}
\usepackage{pb-diagram,pb-xy}

\input{def.tex}

\begin{document}

\thispagestyle{empty}

\noindent
ULM-TP/02-3\\
April 2002\\

\vspace*{1cm}

\begin{center}

{\LARGE\bf   A semiclassical Egorov theorem and \\
\vspace*{1mm} quantum ergodicity \\ 
\vspace*{3mm} for matrix valued operators} \\
\vspace*{3cm}
{\large Jens Bolte}%
\footnote{E-mail address: {\tt jens.bolte@physik.uni-ulm.de}}
{\large and Rainer Glaser}%
\footnote{E-mail address: {\tt rainer.glaser@physik.uni-ulm.de}}

\vspace*{1cm}

Abteilung Theoretische Physik\\
Universit\"at Ulm, Albert-Einstein-Allee 11\\
D-89069 Ulm, Germany 
\end{center}

\vfill

\begin{abstract}
We study the semiclassical time evolution of observables given by matrix
valued pseudodifferential operators and construct a decomposition of the
Hilbert space $L^2(\rz^d)\otimes\kz^n$ into a finite number of almost
invariant subspaces. For a certain class of observables, that is preserved
by the time evolution, we prove an Egorov theorem. We then associate with
each almost invariant subspace of $L^2(\rz^d)\otimes\kz^n$ a classical
system on a product phase space $\TRd\times\cO$, where $\cO$ is a compact
symplectic manifold on which the classical counterpart of the matrix
degrees of freedom is represented. For the projections of eigenvectors
of the quantum Hamiltonian to the almost invariant subspaces we finally
prove quantum ergodicity to hold, if the associated classical systems are
ergodic.
\end{abstract}

\newpage

\input{intro.tex}
\input{1sec.tex}

\input{2sec.tex}

\input{3sec.tex}

\input{4sec.tex}

\input{5sec.tex}
\input{6sec.tex}

\subsection*{Acknowledgment}
We would like to thank M.~Klein for drawing our attention to the paper
\cite{Sim80}. Financial support by the Deutsche Forschungsgemeinschaft (DFG) 
under contract no. Ste 241/15-1 is gratefully acknowledged.

\vspace*{0.5cm}

\section*{Appendices}
\begin{appendix}
\input{app.poisson.tex}
\input{app.skewprod.tex}
\end{appendix}

{\small
\bibliographystyle{amsalpha}
\bibliography{literatur}} 

\end{document}

%% file: def.tex
\newcommand{\ud}{\mathrm{d}}
\newcommand{\ui}{\mathrm{i}}
\newcommand{\ue}{\mathrm{e}}

\newcommand{\dl}{\mathrm{d}\ell}
\newcommand{\TRd}{\mathrm{T}^*\rz^d}
\newcommand{\skl}{\mathrm{S}}
\newcommand{\scl}{\mathrm{S}_{\mathrm{cl}}}
\newcommand{\sinv}{\mathrm{S}_{\mathrm{inv}}}
\newcommand{\SU}{\mathrm{SU}}
\newcommand{\U}{\mathrm{U}}
\newcommand{\su}{\mathrm{su}}
\newcommand{\uu}{\mathrm{u}}

\newcommand{\al}{\alpha}
\newcommand{\be}{\beta}
\newcommand{\de}{\delta}
\newcommand{\De}{\Delta}

\newcommand{\ga}{\gamma}
\newcommand{\Ga}{\Gamma}
\newcommand{\la}{\lambda}

\newcommand{\Om}{\Omega}
\newcommand{\si}{\sigma}

\newcommand{\ve}{\varepsilon}

\newcommand{\cB}{{\mathcal B}}
\newcommand{\cH}{{\mathcal H}}
\newcommand{\cL}{{\mathcal L}}
\newcommand{\cP}{{\mathcal P}} 
\newcommand{\tcP}{\tilde {\mathcal P}} 
\newcommand{\cS}{{\mathcal S}}
\newcommand{\cU}{{\mathcal U}}
\newcommand{\cO}{{\mathcal O}}
\newcommand{\cQ}{{\mathcal Q}}
\newcommand{\cR}{{\mathcal R}}
\newcommand{\cX}{{\mathcal X}}

\newcommand{\gz}{{\mathbb Z}}
\newcommand{\rz}{{\mathbb R}}
\newcommand{\nz}{{\mathbb N}}
\newcommand{\kz}{{\mathbb C}}

\DeclareMathOperator{\tr}{Tr}
\DeclareMathOperator{\mtr}{tr}
\DeclareMathOperator{\vol}{vol}        
\DeclareMathOperator{\supp}{supp}
\DeclareMathOperator{\op}{op}
\DeclareMathOperator{\symb}{symb}
\DeclareMathOperator{\ran}{ran}
\DeclareMathOperator{\id}{id}
\DeclareMathOperator{\Ad}{Ad}
\DeclareMathOperator{\ad}{ad}
\DeclareMathOperator{\ts}{T}
\DeclareMathOperator{\aut}{aut}
\DeclareMathOperator{\inn}{I}

\DeclareMathOperator{\spec}{spec}

\newcommand{\dpr}{\partial}  
\newcommand{\mat}{\mathrm{M}} 

\newcommand{\mtiny}[1]{\text{\tiny\ensuremath{#1}}}  

\newcommand{\eins}{\mathds{1}}

\numberwithin{equation}{section}

\newtheorem{theorem}{Theorem}[section]
\newtheorem{lemma}[theorem]{Lemma}
\newtheorem{prop}[theorem]{Proposition}
\newtheorem{cor}[theorem]{Corollary}

\theoremstyle{definition}
\newtheorem{defn}[theorem]{Definition}
\newtheorem{rem}[theorem]{Remark}

%% file: intro.tex
\section*{Introduction}
\label{sec:intro}
The relation between dynamical properties of a quantum system and its
classical limit is a central subject in the field of quantum chaos.
In this context quantum ergodicity is a well-established concept 
\cite{Zel87,Col85,HelMarRob87,Zel96}. It states for quantisations of ergodic
classical systems that the phase space lifts of almost all eigenfunctions 
of the quantum Hamiltonian converge in the semiclassical limit to an 
equidistribution on the level surfaces of the classical Hamiltonian.
The principal goal of this paper is to establish quantum ergodicity in
systems whose degrees of freedom can be divided into two classes such that 
they are represented in the Hilbert space $L^2(\rz^d) \otimes \kz^n$.
The semiclassical limit shall be performed in terms of a parameter 
$\hbar\to 0$ which is primarily linked to the (translational) degrees of 
freedom that are described by the infinite-dimensional factor $L^2(\rz^d)$. 
The finite dimension $n$ of the other factor is fixed. Examples for systems 
where this description can be applied are relativistic particles with spin 
$1/2$ in slowly varying external fields governed by a Dirac-Hamiltonian, 
or adiabatic situations modelled with a Born-Oppenheimer Hamiltonian. 

This setting leads to a representation of quantum mechanical observables as 
matrix valued pseudodifferential operators acting on 
$L^2(\rz^d) \otimes \kz^n$, whose symbols are suitable matrix valued 
functions on the phase space $\TRd =\rz^d\times\rz^d$ 
with an expansion in $\hbar$. In particular, the principal symbol $H_0$ of 
the selfadjoint quantum Hamiltonian $\cH$ is a hermitian matrix valued 
function on $\TRd$. Its spectral resolution requires to introduce several 
classical dynamics on $\TRd$, each of them generated by one eigenvalue of 
$H_0$. Lifted to the quantum level, this structure results in a decomposition 
of the Hilbert space $L^2(\rz^d)\otimes\kz^n$ into almost invariant 
subspaces with respect to the dynamics generated by the quantum Hamiltonian 
$\cH$ that is directly associated with the spectral resolution of $H_0$.
Recently the case of matrix valued operators for certain quantum 
Hamiltonians with scalar principal symbol, such that on the classical 
side one still has to deal with a single system, has been considered in 
\cite{BolGla00,BolGlaKep01}. Here we extend this approach to the general 
setting of matrix valued operators where one has to define suitable 
classical systems corresponding to each almost invariant subspace of the 
Hilbert space.

So far it appears that the semiclassical limit has only been performed
with respect to one type of the degrees of freedom. For a complete 
(semi-)classical description of the quantum systems under consideration 
one would also require the second type of degrees of freedom, that are 
represented by the factor $\kz^n$ of the Hilbert space 
$L^2(\rz^d) \otimes \kz^n$, to be transferred to a classical level. It 
turns out, however, that for this purpose no further semiclassical parameter 
is needed and the dimension $n$ of the second factor can be held fixed. 
Indeed, a suitable Stratonovich-Weyl calculus \cite{Str57} 
allows to map the principal symbols (with respect to the parameter $\hbar$) 
of observables and their dynamics in a one-to-one manner to genuinely 
classical systems associated with the decomposition of 
$L^2(\rz^d)\otimes\kz^n$ into almost invariant subspaces. On this classical 
level the hierarchy of the two types of degrees of freedom is reflected in 
the structure of the classical dynamics: these are skew-product flows built 
over the Hamiltonian dynamics generated by the eigenvalues of $H_0$.      
 
Apart from classical ergodicity the proof of quantum ergodicity typically
requires two essential inputs. The first one is a suitable version of an 
Egorov theorem \cite{Ego69} that allows to express the time evolution of 
quantum observables in the semiclassical limit in terms of a classical 
dynamics of principal symbols. We achieve this in two steps: beginning 
with matrix valued principal symbols, we proceed to a completely
classical level by exploiting the Stratonovich-Weyl calculus 
in the form developed in \cite{FigGraVar90}. It is
in the last step where the skew-product flows become relevant. The second
input is a Szeg\"o-type limit formula that relates averaged expectation
values of observables to classical phase space averages. This can be
obtained by a straight-forward generalisation of previous results
\cite{HelMarRob87,BolGla00}.

Our main results are the Egorov theorem in section~\ref{sec:Egorov} and 
the quantum ergodicity theorem in section~\ref{sec:EquiDist}. In order to 
formulate the Egorov theorem we first identify a subalgebra in the class 
of bounded semiclassical pseudodifferential operators that is invariant 
under the time evolution. The operators in this subalgebra have 
to be block-diagonal with respect to the projections onto the almost 
invariant subspaces of $L^2(\rz^d) \otimes \kz^n$. Theorem~\ref{thm:Egorov} 
then asserts that the (matrix valued) principal symbol of each block is 
evolved with the Hamiltonian flow associated with that block. In addition, 
it is conjugated with unitary transport matrices that describe the time 
evolution of the matrix degrees of freedom along the trajectories of the
Hamiltonian flow. 

We next identify the dynamics provided by the transport matrices with a
coadjoint action of a certain Lie group. Kirillov's method of orbits
\cite{Kir76} then enables us to connect the apparently quantum mechanical
dynamics with a genuinely classical dynamics on a certain coadjoint orbit
$\cO$, which is a symplectic manifold. This relation can be constructed 
explicitly with the help of the Stratonovich-Weyl calculus developed in 
\cite{FigGraVar90}. As a result we obtain that after a Stratonovich-Weyl
transform the principal symbol of each block of an observable is evolved 
with a skew-product dynamics on the combined symplectic phase space 
$\TRd\times\cO$. This observation restores the general picture behind
Egorov-type theorems: in leading semiclassical order the quantum mechanical
time evolution is determined by classical dynamics.

The decomposition of $L^2(\rz^d) \otimes \kz^n$ into almost invariant 
subspaces and the corresponding set of classical flows force quantum 
ergodicity to be concerned with projections of the eigenvectors
of $\cH$ to the subspaces, since only these are associated with unique
classical systems. The projected eigenvectors, however, are no longer
genuine eigenvectors of $\cH$, but only provide approximate solutions
to the eigenvalue problem and thus yield, after normalisation, quasimodes 
(see \cite{Laz93}). For the latter we prove quantum ergodicity to hold 
in the usual sense. In this context the relevant version of the Egorov 
theorem introduces on the classical side the skew-product flow associated
with the given subspace as described above. We show that if this flow 
is ergodic, the phase space lifts of almost all normalised projected 
eigenvectors converge to equidistribution on the product phase space.
 

%% file: 1sec.tex
\section{Background on matrix valued pseudodifferential \\ operators}
\label{sec:MatixPsiDO}
In this section we recall some basic results of pseudodifferential calculus 
which are well known in the context of operators with scalar symbols. They 
carry over to the case of matrix valued symbols by only slight modifications 
of the results known for operators with scalar symbols which can, e.g., be 
found in \cite{Rob87,DimSjo99}; for the matrix valued case see also
\cite{BolGla00}.

The quantities we are primarily concerned with are linear and continuous 
operators 
$\cB: \mathscr{S}(\rz^d) \otimes \kz^n  \to \mathscr{S}'(\rz^d)\otimes\kz^n$ 
with Schwartz kernels $K_\cB$ taking values in the $n \times n$ matrices 
$\mat_n(\kz)$. Instead of using a kernel 
$K_\cB\in\mathscr{S}'(\rz^d \times \rz^d) \otimes \mat_n(\kz)$ an operator 
$\cB$ can alternatively be represented by its (Weyl) symbol
$B \in \mathscr{S}'(\TRd) \otimes \mat_n(\kz)$ that is related to the 
Schwartz kernel through
\begin{equation} 
\label{eq:WeylKern}
 K_\cB(x,y) = \frac{1}{(2 \pi \hbar)^d} \int_{\rz^d} 
 \ue^{\frac{\ui}{\hbar}(x-y) \xi} B \Bigl( \frac{x+y}{2}, 
 \xi \Bigr) \ \ud \xi.
\end{equation}
Here $\hbar \in (0,\hbar_0]$, with $\hbar_0>0$, serves as a semiclassical 
parameter and $\TRd := \rz^d \times \rz^d$ denotes the cotangent bundle 
of the configuration space $\rz^d$, i.e., $\TRd$ is the phase space of the 
translational degrees of freedom. Below (see section \ref{sec:eigendyn}) 
$\TRd$ will provide one component of a certain combined phase space, which 
also represents the degrees of freedom described by the matrix character of 
the symbol in terms of points on a suitable symplectic manifold.

According to the Schwartz kernel theorem every continuous linear map 
$\cB:\mathscr{S}(\rz^d)\otimes\kz^n \to \mathscr{S}'(\rz^d)\otimes\kz^n$ can 
be viewed as an operator with kernel of the above form. However, operators 
with kernels in $\mathscr{S}'(\TRd) \otimes \mat_n(\kz)$ are too general for 
many purposes; e.g., they can in general not be composed with each other. One 
therefore has to restrict to smaller sets of kernels and hence to smaller 
classes of symbols. To achieve this we make use of order functions 
$m:\TRd \to (1,\infty)$, which have to fulfill a certain growth property in 
the sense that there are positive constants $C$, $N$ such that
\begin{equation*} 
 m(x,\xi) \le C \left( 1+ (x-y)^2 + (\xi-\eta)^2 \right)^{N/2} m(y,\eta)
\end{equation*}
for all $(x,\xi),(y,\eta) \in \TRd$. A typical example for such an order 
function is given by
\begin{equation*}
 m(x,\xi) = \bigl( 1 +x^2 + \xi^2 \bigr)^M,\quad M \ge 0.
\end{equation*}
This notion allows us to define the symbol classes which we will employ 
in the subsequent discussions (see \cite{DimSjo99}).
\begin{defn} 
\label{def:symbols}
Let $m:\, \TRd \rightarrow (1,\infty)$ be an order function. Then define the
symbol class $\skl(m) \subset C^\infty(\TRd)\otimes\mat_n(\kz)$ to be the 
set of $B \in C^\infty(\TRd)\otimes\mat_n(\kz)$ such that for every 
$(x,\xi) \in \TRd$ and all $\alpha,\beta \in \nz_0^d$ there exist constants 
$C_{\alpha,\beta} >0$ with
\begin{equation} 
\label{eq:SklSymb}
 \| \dpr_\xi^\alpha \dpr_x^\beta B(x,\xi) \|_{n \times n} 
 \le C_{\alpha,\beta} m(x,\xi).
\end{equation}
Here $\| \cdot \|_{n \times n}$ denotes an arbitrary (matrix) norm on 
$\mat_n(\kz)$. If in  addition the symbol $B(x,\xi;\hbar)$ depends on 
the parameter $\hbar \in (0,\hbar_0]$, we say that $B \in \skl(m)$ if 
$B(\cdot,\cdot;\hbar)$ is uniformly bounded in $\skl(m)$ when $\hbar$ varies 
in $(0,\hbar_0]$. In particular, for $q \in \rz$ let $\skl^q(m)$ consist of 
$B: \TRd\times (0,\hbar_0] \to \mat_n(\kz)$ belonging to $\hbar^{-q} \skl(m)$, 
i.e.,
\begin{equation*}
 \| \dpr_\xi^\alpha \dpr_x^\beta B(x,\xi;\hbar) \|_{n \times n} 
 \le C_{\alpha,\beta} \hbar^{-q} m(x,\xi)
\end{equation*}
for all $\alpha,\beta \in \nz_0^d$, $(x,\xi) \in \TRd$, and $\hbar 
\in (0,\hbar_0]$.

An asymptotic expansion of $B \in \skl^{q_0}(m)$ is defined by a sequence 
$\{ B_j \in \skl^{q_j}(m) \}_{j \in \nz_0}$ of symbols, where $q_j$ decreases 
monotonically to $-\infty$ and
\begin{equation*}
 B- \sum_{j=0}^N B_j\in \skl^{\scriptstyle q_{_{N+1}}}(m)
\end{equation*}
for all $N \in \nz_0$. In this case we write 
\begin{equation*}
 B \sim \sum_{j=0}^\infty B_j.
\end{equation*}
In the following we will often use the class $\scl^q(m)$ of classical symbols,
whose elements have asymptotic expansions in integer powers of $\hbar$,
\begin{equation*}
 B \sim \sum_{j=0}^\infty \hbar^{-q+j} B_j,
\end{equation*}
where $B_j \in \skl(m)$ is independent of $\hbar$. We also use the notation
\begin{equation*}
 \skl^\infty(m) := \bigcup_{q\in\rz} \skl^q(m) \quad\text{and}\quad
 \skl^{-\infty}(m) := \bigcap_{q\in\rz} \skl^q(m).
 \end{equation*}
\end{defn}
An operator with a kernel of the form (\ref{eq:WeylKern}) and symbol 
$B \in \skl(m)$ clearly maps both $\mathscr{S}(\rz^d) \otimes \kz^n$ and
$\mathscr{S}'(\rz^d) \otimes \kz^n$ into themselves, whereby according to 
(\ref{eq:WeylKern}) it acts on $\kz^n$-valued functions 
$\psi\in\mathscr{S}(\rz^d)\otimes\kz^n$ as
\begin{equation*} 
 (\cB\psi)(x) = \bigl(\op^W[B]\psi\bigr)(x) = \frac{1}{(2 \pi \hbar)^d} 
 \iint_{\TRd} \ue^{\frac{\ui}{\hbar} (x-y) \xi} B \Bigl( 
 \frac{x+y}{2},\xi \Bigr) \psi(y) \ \ud y \, \ud \xi.
\end{equation*}
Operators $\cB=\op^W[B]$ of this type are called Weyl operators, and
$\symb^W[\cB]=B$ denotes the Weyl symbol of $\cB$. If the Weyl symbol of an
operator is a classical symbol with asymptotic expansion 
$B \sim \sum_{j \in \nz_0} \hbar^{-q+j} B_j$, we also call $\op^W[B]$ a 
semiclassical pseudodifferential operator. The leading order term 
$\symb^W_P[\cB]=B_0$ is then referred to as the principal symbol of $\cB$, 
and the subsequent term $B_1$ as the subprincipal symbol. 

The set of Weyl operators with symbols from the classes $\skl(m)$ is stable 
under operator multiplication, in the sense that the operator product is 
again a Weyl operator with symbol in a certain class:
\begin{lemma} 
\label{lem:CompForm}
Let $m_1,m_2$ be order functions. Then for $B_j \in \skl(m_j)$, $j=1,2$, 
the product of the corresponding operators $\cB_j=\op^W[B_j]$ is again a
Weyl operator that can be expressed in terms of the symbols $B_1,B_2$ as
\begin{equation*}
 \cB_1 \cB_2 = \op^W[B_1] \op^W[B_2] = \op^W[B_1 \# B_2],
\end{equation*}
where the symbol product $(B_1,B_2) \mapsto B_1 \# B_2$ is 
continuous from $\skl(m_1) \times \skl(m_2)$ to $\skl(m_1 m_2)$
in the topology generated by the seminorms associated with 
the estimate (\ref{eq:SklSymb}). In explicit terms the symbol product reads
\begin{equation*} 
 (B_1 \# B_2)(x,\xi) = \left. \ue^{\frac{\ui \hbar}{2} \sigma(\dpr_x,\dpr_\xi;
 \dpr_y, \dpr_\eta)} 
  B_1(x,\xi) B_2(y,\eta) \right|_{\substack{y=x\\ \eta=\xi}},
\end{equation*}
where $\sigma(v_x,v_\xi;w_x,w_\xi)=v_x \cdot w_\xi - v_\xi \cdot w_x$
denotes the symplectic two-form on $\TRd$. Furthermore, $B_j\in\scl^0(m_j)$ 
are mapped to $B_1 \# B_2 \in\scl^0(m_1m_2)$ with (classical) asymptotic 
expansion 
\begin{equation*} 
 (B_1 \# B_2)(x,\xi) \sim \sum_{k,j_1,j_2\in \nz_0} \left. 
 \frac{\hbar^{k+j_1+j_2}}{k!} \left( \frac{\ui}{2} 
 \sigma(\dpr_x,\dpr_\xi;\dpr_y,\dpr_\eta)\right)^k B_{1,j_1}(x,\xi) 
 B_{2,j_2}(y,\eta) \right|_{\substack{y=x\\ \eta=\xi}}.
\end{equation*}
\end{lemma}
The following result, which in its original version is due to Beals 
\cite{Bea77}, is useful in situations where one wishes to identify a given 
operator as a pseudodifferential operator.
\begin{lemma} 
\label{lem:Beals}
Let $\cB(\hbar):\mathscr{S}(\rz^d)\otimes\kz^n \to \mathscr{S}'(\rz^d) 
\otimes\kz^n$ be a linear and continuous operator depending on the 
semiclassical parameter $\hbar \in (0,\hbar_0]$. The following statements are 
then equivalent:
\begin{itemize}
\item[(i)] $\cB(\hbar)=\op^W[B]$ is a Weyl operator with symbol 
 $B \in \skl^0(1)$. 
\item[(ii)] For every sequence $l_1(x,\xi),\ldots,l_N(x,\xi)$, $N\in\nz$, of 
 linear forms on $\TRd$ the operator given by the multiple commutator 
 $\displaystyle [\op^W[l_N],[\op^W[l_{N-1}], \cdots,[\op^W[l_1],\cB] \cdots]$ 
 is bounded as an operator on $L^2(\rz^d) \otimes \kz^n$ and its norm is
 of the order $\hbar^N$. 
\end{itemize}
\end{lemma}
The direction (i)$\Rightarrow$(ii) is a simple  consequence
of the symbolic calculus outlined above. For the reverse direction see 
\cite{HelSjo88,DimSjo99}.

In the discussions below we will basically encounter two types of (Weyl)
operators: quantum Hamiltonians $\cH=\op^W[H]$ with symbols $H\in\scl^0(m)$
generating the quantum mechanical time evolution, and observables 
$\cB=\op^W[B]$. In typical cases a Hamiltonian $\cH$ is given and one is 
interested in a suitable algebra of observables that allows to study
dynamical properties of the quantum system. For this purpose it is often
convenient to consider bounded operators. In the scalar case it is sufficient 
to know the boundedness of the symbols in order to obtain a bounded 
Weyl operator. This result, originally going back to Calder\'on and 
Vaillancourt \cite{CalVai71}, generalises to the context of 
pseudodifferential operators with matrix valued symbols without changes.
\begin{prop} 
\label{prop:CalderonVaillancourt}
Let $B(\hbar)\in\skl(1)$, then the Weyl quantisation $\op^W[B(\hbar)]$ of 
this symbol is continuous on $L^2(\rz^d) \otimes \kz^n$. Furthermore, for 
$\hbar \in (0,\hbar_0]$ there exists an upper bound for the operator norm 
of $\op^W[B(\hbar)]$.
\end{prop}
For a proof of this result in the context of semiclassical pseudodifferential
operators (depending on a parameter $\hbar$) see 
\cite{Rob87,HelSjo88,DimSjo99}.   

A quantum Hamiltonian is required to be (essentially) selfadjoint. Thus,
in the case of a Weyl operator $\cH=\op^W[H]$ one requires the symbol
$H$ to take values in the hermitian $n\times n$ matrices. In order to
trace back spectral properties of $\cH$ to properties of the principal symbol 
$H_0$ we will have to construct (asymptotic) inverses of $\cH -z$ and
relate them to $(H_0-z)^{-1}$. In this context an operator $\cB=\op^W[B]$
is called elliptic, if its symbol $B \in \skl(m)$ is invertible, i.e., if 
the matrix inverse $B^{-1}$ exists in $\skl(m^{-1})$. In such a case one can 
construct a parametrix $Q \in \skl(m^{-1})$ which is an asymptotic inverse 
of $B$ in the sense of symbol products,
\begin{equation*} 
 B \# Q \sim Q \# B \sim 1.
\end{equation*}
To see that such an inverse exists for elliptic operators, consider
\begin{equation*}
 \op^W[B] \op^W[B^{-1}] = 1- \hbar \op^W[R],
\end{equation*}
with $R \in \skl(m)$. For sufficiently small $\hbar$ the operator 
$1 - \hbar \op^W[R]$ possesses a bounded inverse and one can define a 
(left and right) inverse $\op^W[B^{-1}](1-\hbar \op^W[R])^{-1}$ for 
$\op^W[B]$. Furthermore, the Beals characterisation of pseudodifferential 
operators (Lemma \ref{lem:Beals}) implies that this inverse is again 
a bounded pseudodifferential operator, see also \cite{DimSjo99}. To obtain an 
asymptotic expansion for the parametrix $Q$ one next defines the operator 
$\cQ_N:=\op^W[B^{-1}](1+ \hbar\cR + \cdots + \hbar^N\cR^N)$, with 
$\cR=\op^W[R]$, which is equivalent to $\cQ=\op^W[Q]$ modulo terms of order 
$\hbar^{N+1}$. One can hence write
\begin{equation} 
\label{eq:ParamAsympExp}
 Q \sim B^{-1} + \hbar (B^{-1} \# R) + \hbar^2 (B^{-1} \# R \# R) + \cdots,
\end{equation}
and finally observes:
\begin{lemma} 
\label{lem:Parametrix}
Let $B \in \skl(m)$ be elliptic in the sense that $B^{-1}(x,\xi)$ exists
for all $(x,\xi) \in \TRd$ and is in the class $\skl(m^{-1})$. Then there 
exists a parametrix $Q \in \skl(m^{-1})$ with an asymptotic expansion of 
the form (\ref{eq:ParamAsympExp}) such that
\begin{equation*}
 B \# Q \sim Q \# B \sim 1.
\end{equation*}
\end{lemma}
From (\ref{eq:ParamAsympExp}) one moreover observes that an elliptic operator
with classical symbol has a parametrix that is a classical symbol.

Frequently it is very convenient to have a functional calculus of 
pseudodifferential operators available. In some places, e.g., we would like 
to apply the Helffer-Sj\"ostrand formula, which shows that a smooth and 
compactly supported function $f \in C_0^\infty(\rz)$ of an essentially 
selfadjoint operator $\cB$ with symbol in $B\in \skl(m)$ yields a 
pseudodifferential operator 
\begin{equation*}
 f(\cB) = -\frac{1}{\pi}\int_\kz \dpr_{\overline{z}} \tilde f(z) (\cB-z)^{-1} 
 \ \ud z,
\end{equation*}
whose symbol is in $\skl(m^{-N})$ for every $N \in \nz$. Here 
$\tilde f\in C_0^\infty (\kz)$ is an almost-analytic extension of $f$, i.e.,
$\tilde f(z)=f(z)$ for $z\in\rz$ and $|\overline{\dpr}\tilde f(z)|\le 
C_N |\mathrm{Im}\,z|^N$ for all $N\in\nz_0$. 
In the scalar case these results were shown in 
\cite{HelSjo89} (see also \cite{DimSjo99}) and have been extended to the
matrix valued situation in \cite{Dim93,Dim98}. A criterion that guarantees the 
essential selfadjointness of $\cB$ is that first its symbol $B \in \skl(m)$ 
is hermitian and, second, that $B + \ui \in \skl(m)$ is elliptic in the sense 
described above. If $B \in \scl^0(m)$ one can even write down a classical 
asymptotic expansion for the symbol of the operator $f(\cB)$ whose principal 
symbol reads $f(B_0)$, where $B_0$ denotes the principal symbol of $\cB$, 
see \cite{Rob87,DimSjo99}.

%% file: 2sec.tex
\section{Semiclassical projections} 
\label{sec:projections}
We motivate the following construction of semiclassical projection
operators by considering the time evolution generated by a quantum
Hamiltonian $\cH$, i.e., the Cauchy problem 
\begin{equation} 
\label{eq:gen_CP}
 \ui \hbar \frac{\dpr}{\dpr t} \psi(t) = \cH \psi(t)
\end{equation}
for an essentially selfadjoint operator $\cH$ defined on a dense domain 
$D(\cH)$ in the Hilbert space $L^2(\rz^d) \otimes \kz^n$. If one introduces 
the strongly continuous one-parameter group of unitary operators 
$\cU(t):= \exp( - \frac{\ui}{\hbar} \cH t), \ t \in \rz$, a solution of 
(\ref{eq:gen_CP}) can be obtained by defining $\psi(t):=\cU(t) \psi_0$ for 
$\psi_0 \in D(\cH)$. Therefore the time evolution 
$\cB(t):=\cU(t)^\ast \cB \cU(t)$ of a bounded operator $\cB \in B(L^2(\rz^d) 
\otimes \kz^n)$ in the Heisenberg picture has to fulfill the (Heisenberg)
equation of motion
\begin{equation} 
\label{eq:heisenberg}
 \frac{\dpr}{\dpr t} \cB(t) = \frac{\ui}{\hbar} [\cH,\cB(t)].
\end{equation}
If one assumes $\cB$ and $\cH$ to be semiclassical pseudodifferential operators
with symbols in the classes $\scl^q(1)$ and $\scl^0(m)$, respectively,  
equation (\ref{eq:heisenberg}) yields in leading semiclassical order 
an equation for the principal symbols: 
\begin{equation*}
 \frac{\dpr}{\dpr t} B_0(t) = \frac{\ui}{\hbar} [H_0,B_0(t)] + O(\hbar^0), 
 \quad \hbar \to 0.
\end{equation*}
If one now requires the time evolution to respect the filtration of 
the algebra $\scl^\infty(1) := \bigcup_{q \in \gz} \scl^q(1)$ then,
in particular, the principal symbol $B_0(t)$ should stay in its class that
derives from the associated grading $\scl^q(1)/\scl^{q-1}(1),\ q \in \gz$.
One thus has to restrict to operators whose principal parts $B_0$
commute with the principal symbol $H_0$ of the operator $\cH$. This 
condition is equivalent to a block-diagonal form of $B_0$,
\begin{equation} 
\label{eq:blockdiag_obs}
 B_0(x,\xi)=\sum_{\mu=1}^l P_{\mu,0}(x,\xi) B_0(x,\xi) P_{\mu,0}(x,\xi),
\end{equation}
with respect to the projection matrices $P_{\mu,0}: \TRd \rightarrow
\mat_n(\kz),\ \mu=1,\ldots,l$, onto the eigenspaces corresponding to the
eigenvalue functions $\la_\mu : \TRd \to \rz$ of the hermitian principal 
symbol matrix $H_0: \TRd \rightarrow \mat_n(\kz)$. Since 
(\ref{eq:blockdiag_obs}) is the semiclassical limit of the symbol of the 
operator $\hbar^q \sum_{\mu=1}^l \op^W[P_{\mu,0}] \cB \op^W[P_{\mu,0}]$, 
when $\cB$ 
is a semiclassical Weyl operator with symbol $B \in \scl^q(1)$, one can ask 
how the symbols $P_{\mu,0}$, which are projectors onto the eigenspaces of 
$H_0$ in $\kz^n$, are related to projection operators onto (almost) invariant 
subspaces of $L^2(\rz^d) \otimes \kz^n$ with respect to $\cH=\op^W[H]$. 
We are hence looking for quantisations $\tcP_\mu$ of symbols 
$P_\mu \in \scl^0(1)$, with principal symbols $P_{\mu,0}$, which are 
(almost) orthogonal projections, i.e.,
\begin{equation} 
\label{eq:proj_prop}
 \tcP_\mu \tcP_\mu = \tcP_\mu = \tcP_{\mu}^{\ast}  \quad \text{mod} \
 O(\hbar^\infty)
\end{equation}
in the operator norm. Moreover, in order
that these operators map to almost invariant subspaces of 
$L^2(\rz^d) \otimes \kz^n$ with respect to the time evolution 
$\cU(t)=\exp( -\frac{\ui}{\hbar} \cH t)$ generated by $\cH$, we require them 
to fulfill
\begin{equation} 
\label{eq:proj_comm}
 \|[ \cH, \tcP_\mu] \|_{L^2(\rz^d) \otimes \kz^n} = 0 \quad 
\text{mod} \ O(\hbar^\infty).
\end{equation}
As it will turn out, it is even possible to modify the operators $\tcP_\mu$
in such a way that they satisfy the relation (\ref{eq:proj_prop}) exactly,
i.e., not only mod $O(\hbar^\infty)$.
 
The above requirements lead us to consider (formal) asymptotic expansions 
for the symbols $P_\mu$,
\begin{equation*}
 P_\mu(x,\xi) \sim \sum_{j=0}^\infty \hbar^j P_{\mu,j}(x,\xi),
\end{equation*}
which satisfy (\ref{eq:proj_prop}) and (\ref{eq:proj_comm}) on a (formal)
symbol level:
\begin{equation} 
\label{eq:SymProjProp}
 P_\mu \# P_\mu \sim P_\mu \sim P^\ast_\mu,
\end{equation}
and
\begin{equation} 
\label{eq:SymProjComm}
 [P_\mu,H]_\#:=P_\mu \# H - H \# P_\mu \sim 0.
\end{equation} 
The solutions of the above equations will be called semiclassical 
projections and can be constructed by two different methods. The first one 
is based on solving the recursive problem arising from (\ref{eq:SymProjProp}) 
and (\ref{eq:SymProjComm}) by employing asymptotic expansions of $P_\mu$
and $H$ in $\scl^0(1)$ and $\scl^0(m)$, respectively, using the symbolic 
calculus outlined in section \ref{sec:MatixPsiDO} and finally equating equal 
powers of the semiclassical parameter $\hbar$. For this procedure cf.
\cite{EmmWei96,BruNou99}. The second method employs the Riesz projection 
formula in the context of pseudodifferential calculus \cite{HelSjo88,NenSor01}.
In the following we will pursue the latter method. 

To this end we consider the matrix valued hermitian principal symbol 
$H_0 \in \skl(m)$ of the operator $\cH$, and in the following we assume:
\begin{itemize}
\item[(H0)] The (real) eigenvalues $\lambda_\mu,\ \mu=1,\ldots,l$,
of $H_0$ have constant multiplicities $k_1,\ldots,k_l$ and fulfill the 
hyperbolicity condition 
\begin{equation*}
 |\lambda_{\nu}(x,\xi) - \lambda_\mu(x,\xi) | \ge C m(x,\xi) \quad 
 \text{for} \quad \nu \neq \mu \quad \text{and} \quad |x|+|\xi| \ge c.
\end{equation*}
\end{itemize}
This requirement is analogous to a condition imposed in \cite{Cor82} 
on the eigenvalues of the symbol of an operator in a strictly hyperbolic 
system, i.e., where the eigenvalues are non-degenerate. In particular, the
problem of mode conversion that arises from points where multiplicities
of eigenvalues change is avoided. Since the eigenvalues are solutions of 
the algebraic equation
\begin{equation} \label{eq:CharEq}
 \det \bigl( H_0(x,\xi) - \lambda \bigr) = 
 \sum_{\nu=0}^n \eta_\nu(x,\xi) \lambda^\nu =0,
\end{equation}
they are smooth functions on $\TRd$. Moreover, since $H_0$ is supposed 
to be hermitian, the eigenvalues are bounded by the matrix norm of $H_0$. 
Using the smoothness of the eigenvalues and the hyperbolicity condition (H0), 
one obtains:
\begin{prop} 
\label{prop:projections}
Let $H \in \scl^0(m)$ be hermitian and let the hyperbolicity condition 
(H0) be fulfilled. Then there exist symbols $P_\mu \in \scl^0(1)$
with asymptotic expansions
\begin{equation} 
\label{eq:SclProj}
 P_\mu \sim \sum_{j=0}^\infty \hbar^j P_{\mu,j}, \quad 
 \mu=1,\ldots,l,
\end{equation}
that fulfill the conditions (\ref{eq:SymProjProp}) and 
(\ref{eq:SymProjComm}). In particular, the coefficients 
$P_{\mu,j},\ j \in \nz_0$, are unique, i.e., the symbols $P_\mu$ are uniquely 
determined modulo $\skl^{-\infty}(1)$.

Furthermore, the corresponding almost projection operators 
$\tcP_\mu=\op^W[P_\mu]$ provide a semiclassical resolution of the identity 
on $L^2(\rz^d) \otimes \kz^d$,
\begin{equation*} 
 \tcP_1 + \cdots + \tcP_l = \mathrm{id}_{L^2(\rz^d) \otimes \kz^n} 
 \quad \mathrm{mod} \ O(\hbar^\infty).
\end{equation*}
\end{prop}
\begin{proof}
We use the technique of \cite{HelSjo88,NenSor01} and consider the Riesz 
projections
\begin{equation} 
\label{eq:Riesz}
 P_\mu(x,\xi):= \frac{1}{2 \pi \ui} \int_{\Gamma_\mu(x,\xi)} 
 Q(x,\xi,z)\ \ud z,
\end{equation}
where $\Gamma_\mu(x,\xi)$ is a simply closed and positively oriented regular 
curve in the complex plane enclosing the, and only the, eigenvalue 
$\la_\mu(x,\xi)\in\rz$ of $H_0(x,\xi)$. Moreover, $Q(x,\xi,z)$ denotes a 
parametrix for $H-z$, i.e., $(H-z) \# Q \sim  Q \# (H-z) \sim 1$ that will
be constructed below. For technical considerations one may choose the 
contour as $\Gamma_\mu(x,\xi)=\{ \lambda_\mu(x,\xi)+ \rho_\mu(x,\xi)\, 
\ue^{\ui \varphi},\ 0 \le \varphi \le 2 \pi\}$ with 
$0 < c \le \rho_\mu < \frac{1}{2} \min_{\nu \neq \mu} 
\{ |\lambda_\mu- \lambda_\nu|\}$. Since $H_0$ is hermitian with eigenvalues
$\la_\nu$, $\nu=1,\dots,l$, one can estimate the matrix norm of 
$(H_0 -z)^{-1}$ for $z \in \Gamma_\mu(x,\xi)$ as
\begin{equation*}
 \| (H_0(x,\xi) - z)^{-1} \|_{n\times n} \le \frac{C}{\rho_\mu(x,\xi)}.
\end{equation*}
The condition (H0) then allows to choose $\rho_\mu(x,\xi) \ge c m(x,\xi)$,
so that $H_0 - z$ is elliptic for $z \in \Gamma_\mu$. If 
then $\hbar$ is sufficiently small, also $H-z =H_0 -z +O(\hbar)$ is elliptic
and the relation
\begin{equation*}
 \bigl( H-z \bigr) \# \bigl( H_0 -z \bigr)^{-1} = 1-\hbar R
\end{equation*}
enables one to construct a parametrix $Q(x,\xi,z) \in \scl^0(m^{-1})$ for 
$H-z$ with asymptotic expansion
\begin{equation*}
 Q(x,\xi,z) \sim \sum_{j=0}^\infty \hbar^j 
 Q_{j}(x,\xi,z)
\end{equation*}
in the same manner as in (\ref{eq:ParamAsympExp}), see also 
\cite{Rob87,DimSjo99}. Plugging this expansion into (\ref{eq:Riesz}) one 
obtains 
\begin{equation} 
\label{eq:RieszProj}
\begin{split}
 P_\mu(x,\xi) &= \frac{1}{2\pi\ui} \int_{\Gamma_\mu(x,\xi)} 
                 Q(x,\xi,z) \ \ud z \\
              &\sim \sum_{j=0}^\infty \hbar^{j} \frac{1}{2\pi\ui} 
                    \int_{\Gamma_\mu(x,\xi)} Q_{j}(x,\xi,z) \ \ud z =: 
                    \sum_{j=0}^\infty \hbar^j P_{\mu,j}(x,\xi)
\end{split}
\end{equation} 
by using the Borel construction to sum asymptotic series of symbols.
According to the properties of the Riesz integral the symbols $P_\mu$ 
therefore fulfill (\ref{eq:SymProjProp}) and (\ref{eq:SymProjComm}).
Since these equations have unique solutions modulo $O(\hbar^\infty)$ 
\cite{EmmWei96}, the coefficients $P_{\mu,j}$ are unique.

We now consider more general $z\in\kz$, and by inspecting the above 
construction notice that the parametrix $Q(z)$ is well-defined as long as 
$z$ has a sufficiently large distance from the eigenvalues of $H_0$. 
According to equation (\ref{eq:ParamAsympExp}) its asymptotic expansion 
then reads
\begin{equation*} 
 Q(z) \sim (H_0-z)^{-1} + \hbar (H_0-z)^{-1}
 \# R(z) \# (\mathrm{id}_{\kz^n} + \hbar R(z) + \hbar^2 R(z) \# R(z)+ 
 \cdots ).
\end{equation*}
Since 
\begin{equation*}
 R(z) = \frac{1}{\hbar} \left( 1 - (H-z) \# (H_0-z)^{-1} \right),
\end{equation*}
it follows according to the composition formula of Lemma \ref{lem:CompForm}
that $R(z)$ contains a factor $(H_0-z)^{-1}$, and therefore 
the only singularities of $Q(z)$ are caused by the eigenvalues of $H_0$. 
Thus, according to the Cauchy formula the expression
\begin{equation*}
 P_1 + \cdots + P_l = \frac{1}{2 \pi \ui} \int_{\bigcup_{\mu=1}^l \Gamma_\mu}
 Q(z) \ \ud z 
\end{equation*}
can be replaced by
\begin{equation*} 
 \frac{1}{2 \pi \ui} \int_{\Gamma(r)} Q(z) \ \ud z,
\end{equation*}
where $\Gamma(r)$ is a contour with minimal distance $r$ from
the origin in $\kz$ that encloses all eigenvalues of $H_0$ while 
keeping a sufficient distance from them. The value of the above integral 
does not depend on the particular choice of $\Gamma(r)$ so that one
can take the limit $r \to \infty$ and hence obtains 
\begin{equation*}
 \lim_{r \to \infty} \frac{1}{2 \pi \ui} \int_{\Gamma(r)}
 Q(z) \ \ud z = \lim_{r \to \infty} \frac{1}{2 \pi \ui} \int_{\Gamma(r)}
 (H_0-z)^{-1}\ \ud z = \mathrm{id}_{\kz^n} \quad \text{mod} \quad 
 O(\hbar^\infty).
\end{equation*}
\end{proof}
The so constructed symbols $P_\mu$ yield semiclassical almost projection 
operators
\begin{equation*}
 \tcP_\mu := \op^W[P_\mu]
\end{equation*}
which according to Proposition \ref{prop:CalderonVaillancourt}
are bounded and obviously satisfy the relations (\ref{eq:proj_prop}) and 
(\ref{eq:proj_comm}). Following \cite{Nen99} one can even construct 
pseudodifferential operators $\cP_\mu$ that are semiclassically equivalent 
to $\tcP_\mu$ in the sense that $\|\tcP_\mu - \cP_\mu \|= O(\hbar^\infty)$, 
and which fulfill (\ref{eq:proj_prop}) exactly. To see this, consider
the operator 
\begin{equation*} 
 \cP_\mu:= \frac{1}{2 \pi \ui} \int_{|z-1|=\frac{1}{2}}
 (\tcP_\mu - z)^{-1} \ \ud z,
\end{equation*}
which is well-defined since the spectrum of $\tcP_\mu$ is concentrated 
near $0$ and $1$. Thus $\cP_\mu$ is an orthogonal projector acting 
on $L^2(\rz^d) \otimes \kz^n$, with 
$\| [\cP_\mu, \cH] \| \le c \| [\tcP_\mu,\cH] \| = O(\hbar^\infty)$. 
Since $\cP_\mu$ is close to $\tcP_\mu$ in operator norm, Beals'
characterisation of pseudodifferential operators (see Lemma \ref{lem:Beals})
yields that $\cP_\mu$ is again a pseudodifferential operator with 
symbol in the class $\skl^0(1)$. This has already been noticed in 
\cite{NenSor01} and follows from the fact that $(\tcP_\mu - z)^{-1}$ for 
$|z-1|=1/2$ is a pseudodifferential operator according to the parametrix 
construction of Lemma \ref{lem:Parametrix}. Having projectors available, 
one can also construct (pseudo\-differential) unitary transformations of 
$L^2(\rz^d) \otimes \kz^n$ which convert $\cH$ by conjugation in an almost
block-diagonal form, see 
\cite{Cor83b,LitFly91,BruRob99,NenSor01,PanSpoTeu02prep}. 
Such unitary transformations are obviously not unique, and since for most 
purposes it suffices to work with the projectors we hence refrain from using 
the unitary operators here. 

In view of the fact that $\cP_\mu$ is an orthogonal projector on the
Hilbert space $L^2(\rz^d) \otimes \kz^n$, one can ask if it is possible to 
satisfy also the relation (\ref{eq:proj_comm}) exactly. In other words, to 
what extent can $\cP_\mu$ be related to a spectral projection of $\cH$? 
(See \cite{HelSjo88,Cor00,Cor01} for examples.) 
We want to illustrate this question in the case where the principal symbol 
$H_0$ of $\cH$ possesses two well-separated eigenvalues $\la_\nu<\la_{\nu+1}$ 
with constant multiplicities $k_\nu$ and $k_{\nu+1}$, respectively, among the 
eigenvalues $\la_1,\ldots,\la_l$. For $l=2$ this is exactly 
the situation that occurs in the case of a Dirac-Hamiltonian that we
will discuss in some detail elsewhere \cite{BolGla01prae}. We also assume 
that there exists $\la\in\rz$ separated from the spectrum $\spec(\cH)$ of 
$\cH$ along with a fixed compact subset $W\subset\TRd$ such that
\begin{equation} 
\label{eq:AsymSepEval}
 \lambda-\lambda_\nu(x,\xi) > C m(x,\xi) \quad \text{and} 
 \quad  \lambda_{\nu+1}(x,\xi)-\lambda > C' m(x,\xi)
\end{equation}
for all $(x,\xi) \in \TRd \setminus W$. It follows from these assumptions 
that for $(x,\xi) \in \TRd \setminus W$ one can  replace the contour 
$\Gamma_{\mtiny{<}}:= \bigcup_{\mu=1}^\nu \Gamma_\mu$ in 
\begin{equation} 
\label{eq:ProjGS}
 P_{\mtiny{<}}(x,\xi) := \sum_{\mu=1}^\nu P_\mu(x,\xi) =
 \frac{1}{2 \pi \ui} \int_{\Gamma_{\mtiny{<}}} 
 Q(x,\xi,z)\ \ud z, 
\end{equation}
see (\ref{eq:Riesz}), by a straight line 
$\Ga_+ :=\{z \in \kz; \ z = \lambda + \ui t, \ t \in \rz \}$ that avoids
the eigenvalues of the principal symbol $H_0$ as well as the spectrum of 
$\cH$. Correspondingly, $\Ga_{\mtiny{>}}:= \bigcup_{\mu=\nu+1}^l \Ga_\mu$  
is deformed into $\Ga_-$ given by $\Ga_+$ with reversed orientation. 
Thus, for $(x,\xi) \in \TRd \setminus W$ 
\begin{equation*} 
 P_{\substack{{\mtiny{<}}\\[-0.5ex] {\mtiny{>}}}}(x,\xi) = 
 \frac{1}{2 \pi \ui} \int_{\Ga_\pm} Q(x,\xi,z) \ \ud z.
\end{equation*}
If this relation held true for all $(x,\xi)\in\TRd$, $\tcP_{\mtiny{<}}
=\op^W [P_{\substack{{\mtiny{<}}}}]$ would be semiclassically equivalent to 
the spectral projection of $\cH$ onto the interval $(-\infty,\la)$ given by 
\begin{equation*} 
 \eins_{(-\infty,\la)}(\cH) = \frac{1}{2 \pi \ui} \int_{\Ga_+}
 (\cH-z)^{-1}\ \ud z,
\end{equation*} 
whereas $\tcP_>$ would correspond to $\eins_{(\la,\infty)}(\cH)$. For 
$(x,\xi) \in W$, however, it might happen that $\Ga_\pm$ crosses an eigenvalue 
of $H_0$. But the 
contribution to $P_{\substack{{\mtiny{<}} \\[-0.5ex] {\mtiny{>}}}}(x,\xi)$ 
coming from the region $W$, where the eigenvalue functions have no 
sufficient distance from $\lambda$, can be shown to be semiclassically small. 
Therefore:
\begin{prop} 
\label{prop:SpecSclProj} 
If the eigenvalues $\lambda_1,\ldots,\lambda_l$ of the principal symbol $H_0$ 
are separated according to (H0) and the condition (\ref{eq:AsymSepEval}) is 
fulfilled, the almost projection 
operators $\tcP_{\substack{{\mtiny{<}} \\[-0.5ex] {\mtiny{>}}}}:=
\op^W [P_{\substack{{\mtiny{<}} \\[-0.5ex] {\mtiny{>}}}}]$, whose 
symbols are defined in (\ref{eq:ProjGS}), can be semiclassically 
identified with the spectral projections $\eins_{(-\infty,\lambda)}(\cH)$
and $\eins_{(\lambda,\infty)}(\cH)$ of the operator $\cH$ to the intervals 
$(-\infty,\lambda)$ and $(\lambda,\infty)$, respectively. This means
\begin{equation*} 
 \| \tcP_{\mtiny{<}} - \eins_{(-\infty,\lambda)}(\cH) \| = O(\hbar^\infty)
 \quad\text{and}\quad \| \tcP_{\mtiny{>}} - \eins_{(\lambda,
 \infty)}(\cH) \| = O(\hbar^\infty).
\end{equation*}
A corresponding statement holds for the related orthogonal projectors
$\cP_{\substack{{\mtiny{<}} \\[-0.5ex] {\mtiny{>}}}}$, 
\begin{equation*} 
 \| \cP_{\mtiny{<}} - \eins_{(-\infty,\lambda)}(\cH) \| = O(\hbar^\infty)
 \quad\text{and}\quad \| \cP_{\mtiny{>}} - \eins_{(\lambda,
 \infty)}(\cH) \| = O(\hbar^\infty).
\end{equation*} 
\end{prop}
\begin{proof}
We use that fact that the contour of integration in the definition of 
$P_{\substack{{\mtiny{<}} \\[-0.5ex] {\mtiny{>}}}}$ in (\ref{eq:ProjGS}) can 
be deformed into $\Gamma_\pm$ outside the compact region $W \subset \TRd$. To 
cut off the region $W$ we choose a smooth and compactly supported function 
$\chi \in C_0^\infty(\TRd)$ equal to one on $W$ and use the corresponding 
partition of unity, $1=\chi + (1-\chi)$, to write
\begin{equation*}
 (2 \pi \ui) P_{\substack{{\mtiny{<}} \\[-0.5ex] {\mtiny{>}}}}(x,\xi) 
 \sim \chi(x,\xi) \# 
 \int_{\Gamma_{\substack{{\mtiny{<}} \\[-0.5ex] {\mtiny{>}}}}}
 Q(x,\xi,z) \ud z + (1- \chi(x,\xi)) \# \int_{\Gamma_{\substack{{\mtiny{<}} 
 \\[-0.5ex]  {\mtiny{>}}}}} Q(x,\xi,z) \ \ud z.
\end{equation*}
In the second contribution, whose support is contained in 
$\TRd\setminus W$, one can replace $\Ga_{\substack{{\mtiny{<}}\\[-0.5ex] 
{\mtiny{>}}}}$ by $\Gamma_\pm$. We are thus left with the first term, which 
represents a symbol $p(x,\xi)$ in $\scl^0(1)$ with compact support 
$\supp p \subset \supp \chi$. Here we apply a translation on $\TRd$ mapping 
$p(x,\xi)$ to $\tilde p(x,\xi):=p(x-x_0,\xi-\xi_0)$, with $x_0$ and $\xi_0$ 
chosen such that $\xi=0$ is no longer contained in the support of $\tilde p$, 
i.e., $\tilde p(x,0)=0$ for all $x \in \TRd$. Using the fact that 
the Weyl operators corresponding to $p$ and $\tilde p$ are unitarily
equivalent, see \cite{Fol89}, we therefore obtain
\begin{equation*}
 \| \op^W[p] \| = \| \op^W[\tilde p] \|.
\end{equation*}
In order to estimate $\op^W[\tilde p]$ in operator norm, we  consider
its action on a function $\psi \in \mathscr{S}(\rz^d)\otimes\kz^n$
given by
\begin{equation*}
 \bigl(\op^W[\tilde p] \psi\bigr)(x) = \frac{1}{(2 \pi \hbar)^d} \iint 
 \ue^{\frac{\ui}{\hbar} (x-y) \xi} \tilde p \Bigl( \frac{x+y}{2}, \xi \Bigr) 
 \psi(y) \ \ud y\, \ud \xi.
\end{equation*}
Since the symbol $\tilde p$ vanishes in a neighbourhood of $\xi=0$, one can 
perform an integration by parts after having inserted the operator 
$(\ui \hbar  |\xi|^{-2} (\xi \cdot \dpr_y))^N$, $N \in \nz$. We thus obtain 
that $\op^W[\tilde p]\psi$ vanishes up to terms of order $\hbar^\infty$ 
and, since $\mathscr{S}(\rz^d)\otimes\kz^n$ is dense in 
$L^2(\rz^d)\otimes\kz^n$, conclude
\begin{equation*}
 \| \op^W[\tilde p] \| = \| \op^W[p] \| = O(\hbar^\infty).
\end{equation*}
To finish the proof we now have to estimate 
\begin{equation*}
 \frac{1}{2 \pi \ui} \int_{\Gamma_\pm} (\cH-z)^{-1} \ \ud z 
 - \frac{1}{2 \pi \ui} \int_{\Gamma_\pm} \op^W[(1-\chi) \# Q(x,\xi,z)] \ \ud z
\end{equation*}
in operator norm. To this end consider for $z\in\Ga_\pm$
\begin{equation*}
 \bigl((\cH-z)^{-1} - \op^W[1-\chi] \op^W[Q(x,\xi,z)] \bigr) (\cH-z) 
 = (1- \op^W[1-\chi]) + O(\hbar^\infty) = O(\hbar^\infty),
\end{equation*}
which holds since $\chi$ has compact support, and thus $\op^W[\chi]$ can
be treated in the same manner as $\op^W[p]$ above. At this point the proof is
complete, since for $z \in \Gamma_{\pm}$ the operator $(\cH-z)$ is invertible
and its inverse has a norm that exceeds $O(\hbar^\infty)$.
\end{proof}

%% file: 3sec.tex
\section{Invariant algebra and Egorov theorem} 
\label{sec:Egorov}
In this section our aim is to identify a suitable class of operators that 
is left invariant by the time evolution. Recalling the reasoning from the
beginning of section~\ref{sec:projections}, we are interested in a subalgebra 
of $\scl^\infty(1)$ whose filtration is respected by the time evolution 
generated by the one-parameter group 
$\cU(t)=\exp \left( - \frac{\ui}{\hbar} \cH t \right)$, 
where $\cH$ is an essentially selfadjoint pseudodifferential operator
with symbol $H$ in the class $\scl^0(m)$. The following assumptions on the 
symbol $H$ guarantee the essential selfadjointness of $\cH$ on $\mathscr{S}
(\rz^d) \otimes \kz^n$ (see \cite{DimSjo99}):
\begin{itemize} 
 \item[(H1)] $H \in \scl^0(m)$ is hermitian,
 \item[(H2)] $H_0 + \ui$ is elliptic in the sense that 
   $\|(H_0(x,\xi) + \ui)^{-1}\|_{n\times n} \le c m(x,\xi)^{-1}$. 
\end{itemize}
Under the assumptions (H1) and (H2), $\cU(t)$ therefore defines a 
strongly-continuous unitary one-parameter group.

We now consider the time evolution of an operator 
$\cB \in B(L^2(\rz^d) \otimes \kz^n)$ given by 
\begin{equation} 
\label{eq:TimeEvolvObs}
 \cB(t):= \cU(t)^\ast \cB \cU(t),
\end{equation}
which is, of course, a bounded operator on $L^2(\rz^d) \otimes \kz^n$.
According to Proposition \ref{prop:CalderonVaillancourt} the boundedness of 
$\cB$ is guaranteed by choosing $B \in \scl^q(1)$. Moreover, a conjugation of 
(\ref{eq:TimeEvolvObs}) with 
$\sum_{\mu=1}^l \cP_\mu= \id_{L^2(\rz^d) \otimes \kz^n} + O(\hbar^\infty)$
results in a bounded operator so that
\begin{equation}
\label{eq:time-block}
 \cB(t) = \sum_{\nu,\mu=1}^l \cP_\mu \ue^{\frac{\ui}{\hbar} \cH \cP_\mu t} 
 \cB \ue^{-\frac{\ui}{\hbar} \cH \cP_\nu t} \cP_\nu 
 = \sum_{\nu,\mu=1}^l \ue^{\frac{\ui}{\hbar} \cH \cP_\mu t}
  \cP_\mu \cB \cP_\nu \ue^{-\frac{\ui}{\hbar} \cH \cP_\nu t} \quad \text{mod} 
 \quad O(\hbar^\infty)
\end{equation}
in the operator norm. Here we have used the property $\ue^{-\frac{\ui}{\hbar}
\cH t} \cP_\nu = \ue^{-\frac{\ui}{\hbar}\cH\cP_\nu t}\cP_\nu$ modulo
$O(\hbar^\infty)$ that follows from the Duhamel principle. Now, the 
principal symbol\footnote{We remark that before transferring 
operators to symbol level one can replace $\cP_\mu$ by $\tcP_\mu$ and employ 
the classical asymptotic expansion of the symbol $P_\mu$. This will
only amount to an error of order $\hbar^\infty$.}
of $\cH \cP_\mu$ is a scalar multiple of the identity in the eigenspace 
$P_{\mu,0}\kz^n$ of $H_0$ corresponding to $\lambda_\mu$, i.e., 
$H_0 P_{\mu,0} = \la_\mu P_{\mu,0}$. Thus, for $\mu=\nu$ the operator 
$\exp \left( \frac{\ui}{\hbar} \cH \cP_\mu t \right) \cB \exp \left( 
-\frac{\ui}{\hbar} \cH \cP_\nu t \right)$ 
is a pseudodifferential operator with symbol in the class $\skl^0(1)$, see 
\cite{Ivr98,BolGla00}. But when $\mu \neq \nu$ the corresponding expressions 
are semiclassical Fourier integral operators. In that case the semiclassical 
limit at time $t\neq 0$ is different in nature from that at time zero. 
For a Dirac-Hamiltonian this phenomenon is related to the so-called
``Zitterbewegung'' which we will discuss in more detail in \cite{BolGla01prae}.
Therefore, we are here interested in operators $\cB$ with symbols in 
$B \in \scl^q(1)$ for which $\cU^\ast(t) \cB \cU(t)$ is again a semiclassical
pseudodifferential operator with symbol $B(t) \in \scl^q(1)$. We hence 
introduce the following notion:
\begin{defn} \label{def:DefInvAlg}
A symbol $B \in \scl^q(1)$ is in the invariant subalgebra $\sinv^\infty(1)$ 
of the algebra $\scl^\infty(1)$, if and only if for all finite $t$ the 
(bounded) operator $\cB(t)=\cU^\ast(t)\cB \cU(t)$, $\cB = \op^W[B]$, is a 
semiclassical pseudodifferential operator with symbol $B(t) \in \scl^q(1)$, 
i.e.,
\begin{equation*} 
 \sinv^\infty(1):= \left\{ B \in \scl^q(1) \, ; \, \symb^W[\cU^\ast(t) \cB 
 \cU(t)] \in \scl^q(1) \quad \text{for} \quad t \in [0,T], \ q \in \gz 
 \right\}.
\end{equation*}
\end{defn}
This means that the invariant algebra $\sinv^\infty(1)$ has a filtration,
induced by the filtration of $\scl^\infty(1)$, which is invariant under
conjugation of the corresponding operators with $\cU(t)$. Due to the 
results of \cite{BolGla00} we expect that operators which are block-diagonal 
with respect to the projections $\cP_\mu$ are in the associated 
invariant operator algebra. This statement is made precise in Theorem 
\ref{thm:Egorov} which is a variant of the Egorov theorem \cite{Ego69} 
for general hyperbolic systems. 

Let us first consider an operator $\cB$ with symbol $B\in\scl^q(1)$
that is block-diagonal with respect to the semiclassical projections, i.e.,
\begin{equation*} 
B \sim \sum_{\mu=1}^l P_\mu \# B \# P_\mu \quad \text{in} \quad \scl^q(1).
\end{equation*}
According to the Heisenberg equation of motion (\ref{eq:heisenberg}) its
time evolution $B(t)$ is governed by  
\begin{equation} 
\label{eq:HeisenbergSymb}
 \frac{\dpr}{\dpr t} B(t) \sim 
 \frac{\ui}{\hbar} [H,B(t)]_\#.
\end{equation}
Suppose now that $B(t)$ has a (formal) asymptotic expansion
\begin{equation*} 
B(t) \sim \sum_{j=0}^\infty \hbar^{-q+j} B(t)_j
\end{equation*}
and use the composition formula of Lemma \ref{lem:CompForm} together with the 
fact that the block-diagonal form of an operator $\cB$ is preserved under the 
time evolution, see (\ref{eq:time-block}). On the symbol level the diagonal 
blocks $\cP_\nu\cB(t)\cP_\nu$ then obey the following equation:
\begin{equation*} 
\begin{split} 
 \frac{\dpr}{\dpr t} & \sum_{j=0}^\infty \hbar^{-q+j} B(t)_{\nu \nu,j} \sim \\
 & \sum_{l,j=0}^\infty \sum_{|\alpha|+|\beta| \ge 0}
 \gamma(\alpha,\beta) \hbar^{-q+l+j+|\alpha|+|\beta|-1}
 \left( B(t)_{\nu \nu,l}\,^{(\beta)}_{(\alpha)} \
 H_{\nu,j}\,^{(\alpha)}_{(\beta)}
 - (-1)^{|\alpha|-|\beta|} H_{\nu,j}\,^{(\alpha)}_{(\beta)} 
  B(t)_{\nu \nu,l}\,^{(\beta)}_{(\alpha)} \right).
\end{split}
\end{equation*}
Here we introduced the notation $F^{(\alpha)}_{(\beta)} := \partial^\al_\xi
\partial^\be_x F$ for $F\in C^\infty (\TRd)\otimes\mat_n(\kz)$, as well as
\begin{equation*} 
\begin{split} 
&\ga(\al,\be) := \frac{\ui^{|\al|-|\be|-1}}{2^{|\al|+|\be|}|\al|!|\be|!},\\
&H_\nu := P_\nu \# H \# P_\nu \sim H \# P_\nu \sim \sum_{j=0}^\infty 
          \hbar^j H_{\nu,j}, \\
&B(t)_{\nu\nu} := P_\nu \# B(t) \# P_\nu \sim \sum_{j=0}^\infty 
                  \hbar^{-q+j} B(t)_{\nu\nu,j}.
\end{split}
\end{equation*}
One hence has to solve, by taking $[H_{\nu,0},B(t)_{\nu \nu,0}]=0$ into 
account,
\begin{equation} 
\begin{split} 
\label{eq:recCP}
 [H_{\nu,0},B(t)_{\nu \nu,n+1}] &= - \frac{\dpr}{\dpr t} B(t)_{\nu \nu,n}
 - \frac{1}{2} \Bigl( \{ B(t)_{\nu \nu,n}, H_{\nu,0} \} - \{H_{\nu,0},
 B(t)_{\nu \nu,n} \} \Bigr) - \ui [B(t)_{\nu \nu,n},H_{\nu,1}] \\ & + 
 \sum_{\substack{0 \le l \le n-1 \\ j+|\alpha|+|\beta|=n-l+1}}
 \gamma(\alpha,\beta) \left( B(t)_{\nu \nu,l}\,^{(\beta)}_{(\alpha)} 
 H_{\nu,j}\,^{(\alpha)}_{(\beta)} - (-1)^{|\alpha|-|\beta|} H_{\nu,j}\,
 ^{(\alpha)}_{(\beta)} B(t)_{\nu \nu,l}\, ^{(\beta)}_{(\alpha)} \right).
\end{split} 
\end{equation}
Upon multiplying this commutator equation with the projection matrices 
$P_{\mu,0}$ from both sides one first realises that it is only solvable,
if the diagonal blocks of the right-hand side, that we denote by 
$R_{n,\nu}(t)$, vanish. The off-diagonal blocks on both sides of the
relation (\ref{eq:recCP}) then yield the general structure of the solution,
which reads
\begin{equation} 
\label{eq:CommEqLsg}
 B(t)_{\nu \nu,n+1} = \sum_{\mu=1}^l P_{\mu,0} B(t)_{\nu \nu,n+1}
 P_{\mu,0} + \sum_{\mu \neq \eta} \frac{P_{\mu,0} 
 R_{n,\nu}(t) P_{\eta,0}}{\lambda_\mu
 - \lambda_\eta},
\end{equation}
see also \cite{Cor95}. This demonstrates that one obtains the off-diagonal 
parts of $B(t)_{\nu \nu,n+1}$ with respect to the projection matrices 
$P_{\mu,0}$ from the preceding coefficients of the asymptotic expansion of 
$B(t)_{\nu \nu}$. The diagonal parts then have to be determined by the 
condition that the commutator equation (\ref{eq:recCP}) must possess a 
(non-trivial) solution with initial value $\left. B(t)_{\nu \nu,n+1}\right|_{
t=0} = B_{\nu \nu,n+1}$. Starting with $n=0$, where the sum in (\ref{eq:recCP})
is empty, one has to solve
\begin{equation*}
 P_{\mu,0} \left( \frac{\dpr}{\dpr t} B(t)_{\nu \nu,0}
 + \frac{1}{2} \Bigl( \{ B(t)_{\nu \nu,0}, H_{\nu,0} \} - \{H_{\nu,0},
 B(t)_{\nu \nu,0} \} \Bigr) + \ui [B(t)_{\nu \nu,0},H_{\nu,1}] \right) 
 P_{\mu,0} = 0.
\end{equation*}
Expressions of this type have already been considered in \cite{Spo00},
where it was shown that the above equation is equivalent to 
\begin{equation} 
\label{eq:ModBgl0}
 \frac{\dpr}{\dpr t} \left( P_{\mu,0} B(t)_{\nu \nu,0} P_{\mu,0} \right)
 - \delta_{\nu \mu} \{\lambda_\nu,P_{\mu,0} B(t)_{\nu \nu,0} P_{\mu,0} \}
 - \ui [\tilde H_{\nu \mu,1},P_{\mu,0} B(t)_{\nu \nu,0} P_{\mu,0}]=0,
\end{equation}
see also appendix \ref{app:MatrixPoisson}. Here we have defined the hermitian
$n\times n$ matrix
\begin{equation} 
\label{eq:DefHnumu}
 \tilde H_{\nu \mu,1}:= \ui (-1)^{\delta_{\nu \mu}}
 \frac{\lambda_\nu}{2} P_{\mu,0} \{ P_{\nu,0}, 
 P_{\nu,0} \} P_{\mu,0} - \ui \delta_{\nu \mu}  
  [P_{\nu,0},\{\lambda_\nu, P_{\nu,0}\}] + P_{\mu,0} H_{\nu,1} P_{\mu,0}
\end{equation}
according to (\ref{eq:H1Tilde}) and (\ref{eq:ModBgl}) of appendix 
\ref{app:MatrixPoisson}. Now, equation (\ref{eq:ModBgl0}) is trivially 
fulfilled for $\nu \neq \mu$, and the case $\nu=\mu$ has already been 
considered in \cite{Ivr98,BruNou99,BolGla00}, where it was shown that the 
solution reads
\begin{equation*} 
 B(t)_{\nu \nu,0}(\xi,x) = d_{\nu \nu}^{-1}(x,\xi,t) 
 B_{\nu \nu,0}\bigl(\Phi_\nu^t(x,\xi)\bigl) d_{\nu \nu}(x,\xi,t).
\end{equation*}
In this expression $\Phi_\nu^t:\TRd \rightarrow \TRd$ denotes the Hamiltonian 
flow generated by the eigenvalue $\lambda_\nu$ of $H_0$, and the transport 
matrix $d_{\nu\nu}$ is determined by the equation
\begin{equation} 
\label{eq:TransEq}
 \dot{d}_{\nu \nu}(x,\xi,t) + \ui \tilde 
 H_{\nu \nu,1}\bigl(\Phi^t_\nu(x,\xi)\bigl) d_{\nu \nu}(x,\xi,t) =0,
 \quad d_{\nu \nu}(x,\xi,0) = \mathrm{id}_{\kz^n}.
\end{equation}
One has thus fixed the coefficients $B(t)_{\nu\nu,0}=P_{\nu,0} B(t)_0 
P_{\nu,0}$, i.e., the principal symbol of $\cB(t)$, since the off-diagonal
terms $B(t)_{\nu \mu,0}=P_{\nu,0} B(t)_0 P_{\mu,0}$ vanish and therefore
trivially fulfill (\ref{eq:recCP}). According to (\ref{eq:CommEqLsg}) we 
hence have also determined the off-diagonal parts of the sub-principal
term $B(t)_{\nu \nu,1}$, which vanish as well. The diagonal contributions 
$P_{\mu,0} B(t)_{\nu \nu,1} P_{\mu,0}$ with respect to the projection 
matrices obey $[P_{\eta,0},P_{\mu,0} B(t)_{\nu \nu,1} P_{\mu,0}]=0$ and 
thus can be determined from the relation (\ref{eq:recCP}). As in 
\cite{Ivr98,BolGla00}, we hence obtain a recursive Cauchy problem for the 
coefficients $B(t)_{\nu \nu,n}$ and are now in a position to state:
\begin{theorem} 
\label{thm:Egorov} 
Let $H \in \scl^0(m)$ be hermitian with the property
\begin{equation} 
\label{eq:HamiltonianGrowth}
 \| H_j\,^{(\alpha)}_{(\beta)}(x,\xi) \|_{n\times n} \le C_{\al,\be} \quad 
 \text{for all} \quad (x,\xi) \in \TRd \ \text{and} \ 
 |\alpha|+|\beta|+j \ge 2-\delta_{j0},
\end{equation}
and such that the conditions (H0) and (H2) are fulfilled.
Furthermore, suppose that $B \in \scl^q(1)$ is block-diagonal with respect 
to the semiclassical projections defined in (\ref{eq:SclProj}),
\begin{equation*}  
 B \sim \sum_{\mu=1}^l P_\mu \# B \# P_\mu.
\end{equation*}
Then $B$ is in the invariant algebra $\sinv^\infty(1)$ introduced in 
Definition~\ref{def:DefInvAlg}, i.e., $\cB(t)$ is again a semiclassical 
pseudodifferential operator with symbol $B(t) \in \scl^q(1)$. Furthermore, 
the principal symbol of $\cB(t)$ is given by 
\begin{equation} 
\label{eq:EgorovHS}
 B(t)_0(x,\xi) = \sum_{\nu=1}^l  d_{\nu \nu}^\ast(x,\xi,t) 
 B_{\nu \nu,0}\bigl(\Phi^t_\nu(x,\xi)\bigl) d_{\nu  \nu}(x,\xi,t),
\end{equation}
where $\Phi^t_\nu$ is the Hamiltonian flow generated by the eigenvalue 
$\la_\nu$ of $H_0$, and $d_{\nu \nu}$ is a unitary $n\times n$ matrix which 
is determined by the transport equation (\ref{eq:TransEq}).
\end{theorem}
\begin{proof} 
As in \cite{Ivr98,BolGla00} we start by rewriting (\ref{eq:recCP}) for the
diagonal block of $B(t)_{\nu \nu,n}$ with respect to $P_{\mu,0}$ in the form
\begin{equation} 
\label{eq:BlockCP} 
\begin{split} 
 \frac{\ud}{\ud t} & 
 \left[ d^{-1}_{\nu \mu}(x,\xi,-t) (P_{\mu,0}B(t)_{\nu \nu,n}
 P_{\mu,0})\circ \Phi^{-t \delta_{\nu \mu}}_\nu(x,\xi) d_{\nu \mu}(x,\xi,-t)
 \right] \\ & = \sum_{\substack{0 \le l \le n-1\\j+|\alpha|+|\beta|=n-l+1}}
 \gamma(\alpha,\beta) P_{\mu,0} \left( B(t)_{\nu \nu,l}\, ^{(\beta)}_{(\alpha)}
 H_{\nu,j}\,^{(\alpha)}_{(\beta)} - (-1)^{|\alpha|-|\beta|}
 H_{\nu,j}\,^{(\alpha)}_{(\beta)} B(t)_{\nu \nu,l}\,^{(\beta)}_{(\alpha)} 
 \right) P_{\mu,0},
\end{split} 
\end{equation} 
where $d_{\nu \mu}$ is determined by the transport equation
\begin{equation*}
 \dot{d}_{\nu \mu}(x,\xi,t) + \ui \tilde H_{\nu \mu,1}
 (\Phi^{t \delta_{\nu \mu}}_\nu(x,\xi)) d_{\nu \mu}(x,\xi,t), \quad 
 d_{\nu \mu}(x,\xi,0)= \id_{\kz^n},
\end{equation*}
that generalises (\ref{eq:TransEq}) also to the off-diagonal transport.
And since $\tilde H_{\nu\mu,1}$ is hermitian, the solution $d_{\nu \mu}$
is a unitary $n\times n$ matrix, which in the case $\nu \neq \mu$ is 
obviously given by
\begin{equation*}
 d_{\nu \mu}(x,\xi,t) = \ue^{ - \ui \tilde H_{\nu\mu,1}(x,\xi) t}.
\end{equation*} 
In order to obtain estimates on the derivatives of the symbols
$P_{\mu,0} B(t)_{\nu \nu,n}(t) P_{\mu,0}$ one has to control the behaviour 
of the flow $\Phi^t_\nu$ generated by the eigenvalue $\lambda_\nu$ of $H_0$. 
To this end we first notice that $H_0 \in \skl(m)$ implies the bound 
$| \lambda_\nu(x,\xi) | \le c m(x,\xi)$ on its eigenvalues. Furthermore, 
due to the hyperbolicity condition (H0) the projections 
$P_{\nu,0}$ onto the eigenspaces of $H_0$ are in $\skl(1)$. We then consider 
the first order derivatives ($|\alpha|+|\beta|=1$) of the relation
\begin{equation*}
 H_0(x,\xi) P_{\nu,0}(x,\xi) = \lambda_{\nu}(x,\xi) P_{\nu,0}(x,\xi),
\end{equation*}
which exist since the eigenvalues $\lambda_\nu$ are smooth functions on the 
phase space $\TRd$, see equation (\ref{eq:CharEq}). One thus obtains
\begin{equation*} 
 \lambda_\nu\,^{(\alpha)}_{(\beta)}(x,\xi) P_{\nu,0}(x,\xi) = 
 \bigl( H_0(x,\xi) P_{\nu,0}(x,\xi) \bigr)^{(\alpha)}_{(\beta)}
 - \lambda_\nu(x,\xi) P_{\nu,0}\,^{(\alpha)}_{(\beta)}(x,\xi).
\end{equation*}
Now, since $P_{\nu,0}(x,\xi) P_{\nu,0}\,^{(\alpha)}_{(\beta)}(x,\xi) 
P_{\nu,0}(x,\xi)=0$, a multiplication of the above equation with 
$P_{\nu,0}(x,\xi)$ from both sides yields
\begin{equation}
\label{eq:lambdarel} 
 \lambda_\nu\,^{(\alpha)}_{(\beta)}(x,\xi) P_{\nu,0}(x,\xi) = 
 P_{\nu,0}(x,\xi) H_0(x,\xi)^{(\alpha)}_{(\beta)}P_{\nu,0}(x,\xi), 
\end{equation}
and hence
\begin{equation*} 
 \bigl|\la_\nu\,^{(\al)}_{(\be)}\bigr| = c \bigl\| \la_\nu\,^{(\al)}_{(\be)} 
 P_{\nu,0} \bigr\|_{n\times n} = c \bigl\| P_{\nu,0} H_0\,^{(\al)}_{(\be)}
 P_{\nu,0} \bigr\|_{n\times n}
 \le \tilde c \bigl\| H_0\,^{(\al)}_{(\be)} \bigr\|_{n\times n}.
\end{equation*}
$H_0 \in \skl(m)$ therefore implies that the first order derivatives of 
$\lambda_\nu$ are bounded by the order function $m$. One can continue this 
argument by successively differentiating equation (\ref{eq:lambdarel}), \
and concludes that $\la_\nu \id_{\kz^n} \in \skl(m)$ for all $\nu=1,\ldots,l$.
In particular, the property
\begin{equation*}
 \bigl\| H_0\,^{(\alpha)}_{(\beta)}(x,\xi) \bigr\|_{n\times n} \le 
 C_{\alpha,\beta} \quad \text{for} \quad |\alpha|+|\beta| \ge 1,
\end{equation*}
which follows from (\ref{eq:HamiltonianGrowth}), transfers to a corresponding
growth property of the eigenvalues of $H_0$:
\begin{equation*}
 \bigl| \lambda_\nu\, ^{(\alpha)}_{(\beta)}(x,\xi) \bigr| \le C_{\alpha,\beta} 
 \quad \text{for} \quad |\alpha|+|\beta| \ge 1.
\end{equation*}
Therefore, the Hamiltonian flows $\Phi^t_\nu$ exist globally on $\TRd$ such 
that $|\Phi^t_\nu \, ^{(\alpha)}_{(\beta)}(x,\xi)| \le C_{\alpha,\beta}$ 
for all $\alpha$, $\beta \in \nz_0^d$ and for all finite times $t \in [0,T]$, 
see \cite{Rob87}. This property guarantees that 
$B \circ \Phi_\nu^t \in \skl(1)$ for all $B \in \skl(1)$.
Concerning the unitary matrices $d_{\nu \mu}$  the following is true:
\begin{lemma} 
If the subprincipal symbol $H_1$ of $\cH$ satisfies 
$\| H_1\,^{(\alpha)}_{(\beta)} \|_{n\times n} \le C_{\alpha,\beta}$ 
for all $|\alpha|+|\beta|\ge 1$, then 
$\| d_{\nu \mu} \,^{(\alpha)}_{(\beta)}(x,\xi,t) \|_{n\times n} \le 
C'_{\alpha,\beta}$ for all $t \in [0,T]$, $|\alpha|+|\beta| \ge 1$ 
and $\nu,\, \mu=1,\ldots,l$.
\end{lemma}
For the proof of this lemma see \cite{Ivr98}.
With these properties at hand one can integrate equation (\ref{eq:BlockCP}) 
and solve for $P_{\mu,0} B(t)_{\nu \nu,n} P_{\mu,0}$ by conjugating with 
$d_{\nu \mu}(x,\xi,-t)$ and shifting the arguments by 
$\Phi^{\delta_{\nu \mu} t}_\nu$ (which only amounts to an actual shift in 
the case $\nu=\mu$). For the principal symbol of $\cB(t)$ one thus obtains
\begin{equation*}
 B(t)_{\nu \nu,0}(x,\xi) = d_{\nu \nu}(\Phi^t_\nu(x,\xi),-t)
 B_{\nu \nu,0}(\Phi^t_\nu(x,\xi)) d_{\nu \nu}^{-1}(\Phi^t_\nu(x,\xi),-t),
\end{equation*} 
which is the only block of $B(t)_{\nu \nu,0}$ with respect to 
$P_{\mu,0}$, $\mu=1,\ldots,l$, that is different from zero. Using
\begin{equation} \label{eq:dUnitary}
 d_{\nu \mu}(\Phi^{t \delta_{\nu \mu}}_\nu(x,\xi),-t) = d^{-1}_{\nu \mu}
 (x,\xi,t) = d^\ast_{\nu \mu}(x,\xi,t),
\end{equation}
see \cite{BruNou99}, one finally obtains (\ref{eq:EgorovHS}). For the
higher coefficients $B(t)_{\nu \nu,n}$, $n \ge 1$, one employs the Duhamel
principle and uses that fact that the sum in (\ref{eq:BlockCP}) is taken 
over indices with $|\alpha|+|\beta|+j \ge 2$, and thus involves terms in 
$\skl(1)$, in order to conclude that $B(t)_{\nu \nu,n} \in \skl(1)$.
This shows that one has found an asymptotic expansion in $\scl^q(1)$ for 
the symbol of $\cU^\ast(t) \cB \cU(t)$ that can be summed with the Borel
method to yield a complete symbol.
\end{proof}
This theorem shows that, for finite times $t$, one can associate
to a (semiclassically) block-diagonal symbol $B \in \scl^q(1)$ a symbol 
$B(t) \in \scl^q(1)$ whose quantisation $\op^W[B(t)]$ is semiclassically 
close to $\cB(t)=\cU^\ast(t) \cB \cU(t)$, i.e.,
\begin{equation*}
 \| \cB(t) - \op^W[B(t)] \| = O(\hbar^\infty) \quad 
 \text{for all} \quad t \in [0,T].
\end{equation*}
This is a semiclassical version of the Egorov theorem \cite{Ego69}, which 
was originally formulated for the case of scalar symbols. A weaker 
version that is also sometimes referred to as an Egorov theorem (see, e.g.,
\cite{PanSpoTeu02prep}) would only assert that one can evolve the principal 
symbol $B_0$ of $\cB$ into a symbol $B(t)_0$, as given in (\ref{eq:EgorovHS}),
such that its quantisation $\op^W[B(t)_0]$ is $\hbar$-close to the 
time-evolved operator $\cB(t)$, i.e.,
\begin{equation*}
 \| \cB(t) - \op^W[B(t)_0] \| = O(\hbar).
\end{equation*}
This statement is clearly covered by Theorem \ref{thm:Egorov}, since the 
quantisation of the difference $B(t)-B(t)_0 \in \scl^{q-1}(1)$ yields a 
bounded operator with norm of order $\hbar$, see 
Proposition~\ref{prop:CalderonVaillancourt}.
 
We will now show (generalising results of Cordes \cite{Cor83,Cor00,Cor01})
that the semiclassical block-diagonal operators exhaust all operators with 
symbols in the invariant algebra $\sinv^\infty(1)$.
\begin{prop} 
The invariant subalgebra $\sinv^\infty(1)$ of $\scl^\infty(1)$ consists
of precisely those $B \in \scl^q(1)$ which are semiclassically block-diagonal
with respect to the projections $P_\mu,\, \mu=1,\ldots,l$, defined in 
(\ref{eq:RieszProj}) of Proposition \ref{prop:projections}, i.e.,
\begin{equation*} 
 B \in \sinv^\infty \subset \scl^\infty(1) \quad \Leftrightarrow \quad
 B \sim \sum_{\mu=1}^l P_\mu \# B \# P_\mu.
\end{equation*}
\end{prop}
\begin{proof}
Consider an operator $\cB$ with symbol $B\in\scl^\infty(1)$, whose equation 
of motion is given by (\ref{eq:HeisenbergSymb}). For the symbol of the 
time-evolved operator we now assume an asymptotic expansion
\begin{equation*} 
B(t) \sim \sum_{j=0}^\infty\hbar^{-q+j} B(t)_j 
\end{equation*}
in $\scl^q(1)$. Furthermore, one can use (\ref{eq:SymProjComm}) to separate 
(\ref{eq:HeisenbergSymb}) into blocks with respect to $P_\mu$, 
$\mu=1,\ldots,l$. For the off-diagonal blocks ($\nu \neq \mu$) one 
therefore obtains
\begin{equation} 
\label{eq:HeisenbergND}
 \frac{\dpr}{\dpr t} B(t)_{\nu \mu} \sim \frac{\ui}{\hbar} 
 [H,B(t)_{\nu \mu}]_{\#},
\end{equation}
where $B(t)_{\nu \mu}:= P_\nu \# B(t) \# P_\mu \sim \sum_{j=0}^\infty 
\hbar^{-q+j} B(t)_{\nu \mu,j}$. In leading semiclassical order the factor 
$\hbar^{-1}$ on the right-hand side of equation (\ref{eq:HeisenbergND}) 
enforces the condition
\begin{equation*}
 [H_0,B(t)_{\nu \mu,0}]=(\lambda_\nu - \lambda_\mu)B(t)_{\nu \mu,0}=0.
\end{equation*}
Since $\lambda_\mu \neq \lambda_\nu$ for $\mu \neq \nu$, this immediately 
yields $B(t)_{\nu \mu,0}=0$. Furthermore,
\begin{equation*}
 \frac{\dpr}{\dpr t} \sum_{j=1}^\infty \hbar^{-q+j} B(t)_{\nu \mu,j}
 \sim \ui \Bigl[ H, \sum_{j=1}^\infty \hbar^{-q+j-1} B(t)_{\nu \mu,j} 
 \Bigr]_\#.
\end{equation*}
Again the leading order on the right-hand side has to vanish, i.e.,
\begin{equation*}
 [H_0,B(t)_{\nu \mu,1}] = 0.
\end{equation*}
This means that the symbol $B(t)_{\nu \mu,1}$ must be block-diagonal
with respect to the projection matrices $P_{\mu,0} \in \skl(1)$. But
\begin{equation*}
 P_{\la,0} B(t)_{\nu \mu,1} P_{\la,0} = \symb_P^W \left[ \hbar^{-1}
 \left( P_\la \# (B(t)_{\nu \mu} - B(t)_{\nu \mu,0}) \# P_\la \right) 
 \right] = 0,
\end{equation*}
since $B(t)_{\nu \mu,0}=0$ for $\nu \neq \mu$. Iterating the
above procedure we see that if $B \in \sinv(1)$, then it has to be 
block-diagonal with respect to $P_\mu,\ \mu=1,\ldots,l$. This proves one
direction asserted in the proposition. The other direction, that the 
block-diagonal operators form a subset of the invariant algebra,
is contained in the Egorov theorem \ref{thm:Egorov}.
\end{proof}
At this point we want to add a comment on the transport equation 
(\ref{eq:TransEq}) that not only occurs in connection
with an Egorov theorem, but also in a WKB-type framework. In this context
Littlejohn and Flynn \cite{LitFly91} introduced a splitting of the analogue
to $\tilde H_{\nu\nu,1}$ (defined in equation (\ref{eq:DefHnumu})) into two
contributions, one of which is related to a Berry connection \cite{Ber84}. 
Subsequently Emmrich and Weinstein \cite{EmmWei96} generalised the 
approach of \cite{LitFly91} and gave a geometrical interpretation for
the second contribution, which they related to a Poisson curvature.
We now want to identify the two contributions in the present situation, i.e.,
in $\tilde H_{\nu\nu,1}$. To this end we calculate
$H_{\nu,1}=P_{\nu,1} H_0 + P_{\nu,0} H_1 + \frac{\ui}{2} \{P_{\nu,0}, H_0\}$ 
using
\begin{equation*}
 - P_{\nu,0} P_{\nu,1} P_{\nu,0} + (1- P_{\nu,0}) P_{\nu,1} 
 (1-P_{\nu,0}) = \frac{\ui}{2} \{ P_{\nu,0}, P_{\nu,0} \}, 
\end{equation*}
which follows from the condition $P_\nu \# P_\nu \sim P_\nu$ and the 
composition formula in Lemma \ref{lem:CompForm}. Thus 
\begin{equation*}
 P_{\nu,0} H_{\nu,1} P_{\nu,0} = P_{\nu,0} H_1 P_{\nu,0} 
 + \ui \frac{\lambda_\nu}{2} P_{\nu,0} \{ P_{\nu,0}, P_{\nu,0} \} P_{\nu,0} 
 + \frac{\ui}{2} \sum_{\eta=1}^l \lambda_\eta P_{\nu,0} \{ P_{\nu,0}, 
 P_{\eta,0} \} P_{\nu,0}.
\end{equation*}
The relation $P_{\nu,0} \{ P_{\nu,0}, P_{\eta,0} \} P_{\nu,0} =
-P_{\nu,0} \{P_{\eta,0}, P_{\eta,0} \} P_{\nu,0}$ and the spectral
representation $H_0 =\sum_\mu \la_\mu P_{\mu,0}$ now allow to rewrite
the expression (\ref{eq:DefHnumu}) for $\tilde H_{\nu \nu,1}$ as
\begin{equation*}
 \tilde H_{\nu \nu,1} = H_{\nu,{\rm Berry}} + H_{\nu,{\rm Poisson}} +
 P_{\nu,0} H_1 P_{\nu,0}
\end{equation*}
with
\begin{equation*}
\begin{split}
 H_{\nu,{\rm Berry}}   &:= - \ui [P_{\nu,0},\{\lambda_\nu,P_{\nu,0} \}], \\
 H_{\nu,{\rm Poisson}} &:= \frac{\ui}{2} \Bigl( \la_\nu P_{\nu,0} 
                           \{ P_{\nu,0}, P_{\nu,0} \} P_{\nu,0} 
                           + P_{\nu,0} \{ P_{\nu,0}, H_0 - 
                           \la_\nu P_{\nu,0} \} P_{\nu,0} \Bigr).
\end{split}
\end{equation*}
This corresponds exactly to the splitting discussed in \cite{EmmWei96}, 
see also \cite{Spo00}.

%% file: 4sec.tex
\section{Dynamics in the eigenspaces} 
\label{sec:eigendyn}
According to the Egorov theorem \ref{thm:Egorov}, the semiclassical calculus 
outlined above results not only in a transport of the principal symbols
of observables by the Hamiltonian flows $\Phi_\nu^t$, but also in a 
conjugation by the (unitary $n\times n$) transport matrices $d_{\nu\nu}$.
The latter define the dynamics of those degrees of freedom that on the
quantum mechanical level are described by the factor $\kz^n$ of the total
Hilbert space $L^2(\rz^d)\otimes\kz^n$. Our intention in this section
now is to develop combined classical dynamics of both types of degrees of
freedom, i.e., those described by the Hamiltonian flows and those that are
represented by the conjugations. In this context the fact that the 
conjugations enter along integral curves of the Hamiltonian flows introduces 
a hierarchy among the two types of degrees of freedom. 

In a first step we confirm that the dynamics represented by the transport
matrices $d_{\nu\nu}$ take place in the eigenspaces of the principal
symbol $H_0$ in $\kz^n$. To this end we notice that since at every point 
$(x,\xi) \in \TRd$ the projection matrices $P_{\nu,0}(x,\xi)$ yield an 
orthogonal splitting of $\kz^n$ and have constant rank $k_\nu$, they define 
$k_\nu$-dimensional subbundles $\pi_\nu : E^\nu \rightarrow \TRd$ of the 
trivial vector bundle $\TRd \times \kz^n$ over phase space. The fibre 
$E_{(x,\xi)}^\nu = \pi_\nu^{-1}(x,\xi)$ over $(x,\xi) \in \TRd$ is given 
by the range of the projection, i.e., 
$E_{(x,\xi)}^\nu = P_{\nu,0}(x,\xi) \kz^n$. Furthermore, the canonical 
hermitian structure of $\kz^n$ induces a hermitian structure on the fibres. 
We now intend to interpret the conjugation by $d_{\nu\nu}$ as a dynamics in 
the eigenvector bundle $E^\nu$, and for this purpose notice:
\begin{lemma}
\label{lem:dunitary}
The restricted transport matrices $d_{\nu \nu}(x,\xi,t) P_{\nu,0}(x,\xi)$
provide unitary maps between the fibres $E^\nu_{(x,\xi)}$ and 
$E^\nu_{\Phi^t_\nu(x,\xi)}$.
\end{lemma}
\begin{proof}
In order to see that $d_{\nu \nu}(x,\xi,t) P_{\nu,0}(x,\xi)$ maps 
$E^\nu_{(x,\xi)}$ into $E^\nu_{\Phi^t_\nu(x,\xi)}$ we show 
\begin{equation} \label{eq:SubbdlInv}
 P_{\nu,0}(\Phi^t_\nu(x,\xi)) d_{\nu \nu}(x,\xi,t) P_{\nu,0}(x,\xi) = 
 d_{\nu \nu}(x,\xi,t) P_{\nu,0}(x,\xi).
\end{equation}
This relation is certainly true for $t=0$ where both sides yield $P_{\nu,0}$.
Moreover, the derivative with respect to $t$ of the left-hand side reads
\begin{equation*}
 \{ \lambda_{\nu},P_{\nu,0} \}(\Phi^t_\nu(x,\xi)) d_{\nu \nu}(x,\xi,t) 
 P_{\nu,0}(x,\xi) - \ui  P_{\nu,0}(\Phi^t_\nu(x,\xi)) \tilde H_{\nu \nu,1}
 (\Phi^t_\nu(x,\xi)) d_{\nu \nu}(x,\xi,t) P_{\nu,0}(x,\xi),
\end{equation*}
which equals
\begin{equation*}
 - \ui \tilde H_{\nu \nu,1}(\Phi^t_\nu(x,\xi)) P_{\nu,0}(\Phi^t_\nu(x,\xi))
 d_{\nu \nu}(x,\xi,t) P_{\nu,0}(x,\xi), 
\end{equation*}
since the commutator $[P_{\nu,0},\tilde H_{\nu \nu,1}]$ can be
calculated as (see equation (\ref{eq:DefHnumu}))
\begin{equation*}
 - \ui [P_{\nu,0},[P_{\nu,0},\{\lambda_\nu,P_{\nu,0}\}]] = - \ui \left( 
 P_{\nu,0} 
 \{ \lambda_\nu, P_{\nu,0} \} + \{ \lambda_\nu, P_{\nu,0} \} P_{\nu,0} 
 \right)
 = - \ui \{ \lambda_\nu, P_{\nu,0} \};
\end{equation*}
here we have used (\ref{eq:PlPP}) and $P_{\nu,0}^2=P_{\nu,0}$. 
Thus, $P_{\nu,0}(\Phi^t_\nu(x,\xi)) d_{\nu \nu}(x,\xi,t) P_{\nu,0}(x,\xi)$
fulfills the same differential equation with respect to $t$ as 
$d_{\nu \nu}(x,\xi,t) P_{\nu,0}$, and this finally implies the validity of 
equation (\ref{eq:SubbdlInv}). 

In order to see the unitarity, one has to show that $d_{\nu \nu}(x,\xi,t) 
P_{\nu,0}(x,\xi)$ is an isometry whose range is the complete fibre
$E^\nu_{\Phi^t_\nu(x,\xi)}$. The first point is clear since $d_{\nu \nu}$
is unitary on $\kz^n$ and the fibres inherit their hermitian structures from
$\kz^n$. The second point follows from the observation that the transport
provided by $d_{\nu \nu}$ can be reversed: Given 
$v(\Phi^t_\nu(x,\xi)) \in E_{\Phi^t_\nu(x,\xi)}$, the vector 
$P_{\nu,0}(x,\xi) d_{\nu\nu}(\Phi^t_\nu(x,\xi),-t) v(\Phi^t_\nu(x,\xi))$
lies in $E^\nu_{(x,\xi)}$ and is mapped to $v(\Phi^t_\nu(x,\xi))$ by
$d_{\nu \nu}(x,\xi,t) P_{\nu,0}(x,\xi)$, see (\ref{eq:dUnitary}).
\end{proof}
According to the above, the action of $d_{\nu \nu}(x,\xi,t)$ on a section
in $E^\nu$ can be viewed as a parallel transport along the integral curves
of the flow $\Phi^t_\nu$. If one now introduces sections of $E^\nu$ that 
yield orthonormal bases $\{ e_1(x,\xi), \dots, e_{k_\nu}(x,\xi) \}$ of 
the fibres $E^\nu_{(x,\xi)}$, the representations of $d_{\nu \nu}(x,\xi,t)$ 
in these bases are unitary $k_\nu \times k_\nu$ matrices $D_\nu (x,\xi,t)$. 
Since the principal symbol $H_0$ of the operator $\cH$ is hermitian 
(on $\kz^n$), a preferred choice for the sections $\{ e_1,\dots,e_{k_\nu} \}$ 
would consist of orthonormal eigenvectors of $H_0$. However, this choice is 
obviously not unique because it amounts to fixing an isometry $V_\nu (x,\xi) : 
\kz^{k_\nu} \to E^\nu_{(x,\xi)}$, such that $V_\nu (x,\xi)V_\nu^* (x,\xi) 
= P_{\nu,0}(x,\xi)$ and $V_\nu^* (x,\xi) V_\nu (x,\xi)=\id_{\kz^{k_\nu}}$.
Here one still has a freedom to change the isometry by an arbitrary unitary 
automorphism of $\kz^{k_\nu}$. Having chosen an isometry $V_\nu (x,\xi)$ for 
every fibre $E^\nu_{(x,\xi)}$ in a smooth way, the $n\times n$ transport 
matrices $d_{\nu \nu}(x,\xi,t)$ are mapped to unitary $k_\nu\times k_\nu$ 
matrices
\begin{equation}
\label{eq:DefD}
D_\nu (x,\xi,t) := V_\nu^* \bigl(\Phi_\nu^t(x,\xi)\bigr)d_{\nu \nu}(x,\xi,t)
V_\nu (x,\xi).
\end{equation}
Their dynamics follows from the transport equation (\ref{eq:TransEq}) as
\begin{equation}
\label{eq:DDyn}
\dot{D}_\nu (x,\xi,t) + \ui\tilde H_\nu \bigl(\Phi^t_\nu(x,\xi)\bigr)
D_\nu (x,\xi,t) = 0 \quad \text{with} \quad 
D_\nu (x,\xi,0) = \id_{\kz^{k_\nu}},
\end{equation}
where the hermitian $k_\nu\times k_\nu$ matrix $\tilde H_\nu$ is derived 
from (\ref{eq:DefHnumu}) for $\mu=\nu$,
\begin{equation*} 
\tilde H_\nu = -\ui\frac{\la_\nu}{2}V_\nu^* \{P_{\nu,0},P_{\nu,0}\}V_\nu +
\ui \{\la_\nu,V_\nu^*\}V_\nu + V_\nu^* H_{\nu,1}V_\nu.
\end{equation*}

What is of more importance for later purposes than the non-uniqueness of
this representation, however, is the fact that the above construction allows 
to introduce a skew-product flow over the Hamiltonian flow $\Phi^t_\nu$,
thus reflecting the hierarchy of the two types of degrees of freedom. See
\cite{CorFomSin82} for a definition of  skew-product flows and cf. 
\cite{BolKep99b} where these occur in the context of a semiclassical trace
formula for matrix valued operators. At this stage now provisionally consider 
\begin{equation*}
\hat Y_\nu ^t : \TRd \times \U(k_\nu) \to \TRd \times \U(k_\nu),
\end{equation*}
defined by $\hat Y^t_\nu (x,\xi,g) := ( \Phi^t_\nu (x,\xi),
D_\nu (x,\xi,t)g )$, which yields a flow on the product space 
$\TRd \times \U(k_\nu)$ due to the cocycle relation 
$D_\nu (x,\xi,t+t') = D_\nu (\Phi^t_\nu(x,\xi),t')D_\nu (x,\xi,t)$. 
Later we are interested in ergodic properties of such skew-product flows, 
and these are independent of the particular choice of the sections 
$\{ e_1,\dots,e_{k_\nu} \}$. Here we remark that in some cases the point of 
view advertised above might turn out too general. It can indeed happen that 
the fibre part of the skew-product flow does not require the complete group 
$\U(k_\nu)$. E.g., in \cite{BolGlaKep01} a situation was considered where 
$k_\nu =2j+1,\ j \in \frac{1}{2} \nz$, and the transport matrices $D_\nu$ were
 operators in a $2j+1$-dimensional unitary irreducible representation of 
$\SU(2)$. This fact could be identified by the observation that when $(x,\xi)$
 ranges over $\TRd$, the skew-hermitian matrices $\ui\tilde H_\nu (x,\xi)$ 
generate a Lie subalgebra of $\uu(2j+1)$ which is isomorphic to $\su(2)$.

In the general case one therefore should not necessarily expect that the 
transport matrices $D_\nu$ generate all of $\U(k_\nu)$, but only a certain
Lie subgroup. In order to identify this group we consider the Lie subalgebra 
\begin{equation}
\label{eq:LiegDef}
 \bigl\langle \ui\tilde H_\nu (x,\xi);
\ (x,\xi)\in\TRd \bigr\rangle \subset \uu(k_\nu)
\end{equation}
generated by the skew-hermitian matrices $\ui\tilde H_\nu (x,\xi)$.
Via exponentiation of this subalgebra one hence obtains a Lie subgroup 
$G\subset\U(k_\nu)$ that is compact and connected. To be more precise, the 
result of the exponentiation is a $k_\nu$-dimensional unitary representation 
$\rho$ of $G$. Its Lie algebra $\mathfrak{g}$ then is embedded in 
(\ref{eq:LiegDef}) via the derived representation $\ud\rho$. In this
setting the transport matrices $D_\nu$ are operators in the representation
$\rho$, i.e., $D_\nu (x,\xi,t)= \rho(g_\nu(x,\xi,t))$. Hence we are now in 
a position to define the skew-product flows 
\begin{equation}
\label{eq:skewG}
\tilde Y_\nu ^t : \TRd \times G \to \TRd \times G
\end{equation}
through 
\begin{equation}
\label{eq:skewflowG}
\tilde Y_\nu ^t (x,\xi,g) = \bigl( \Phi^t_\nu(x,\xi),g_\nu(x,\xi,t)g \bigr).
\end{equation}
These flows leave the product measure $\ud x\,\ud\xi\,\ud g$ on $\TRd \times G$
invariant, which consists of Lebesgue measure $\ud x\,\ud\xi$ on $\TRd$
and the normalised Haar measure $\ud g$ on $G$. Moreover, if one restricts
the Hamiltonian flows $\Phi^t_\nu$ to compact level surfaces of the eigenvalue
functions $\la_\nu$ at non-critical values $E$,
\begin{equation*} 
\Om_{\nu,E} := \la_\nu^{-1}(E) = \bigl\{ (x,\xi)\in\TRd;\ 
\la_\nu(x,\xi) =E \bigr\},
\end{equation*}
the restrictions of the skew-product flows $\tilde Y_\nu ^t$ to 
$\Om_{\nu,E}\times G$ leave the measures $\dl(x,\xi)\,\ud g$ invariant, where
$\dl(x,\xi)$ denotes the normalised Liouville measure on $\Om_{\nu,E}$.

Below we are interested in the question under which conditions imposed on
suitable classical dynamics quantum ergodicity holds, see section 
\ref{sec:EquiDist}. In analogy to \cite{BolGla00} one approach to this 
problem would be to consider the restriction of the skew-product flow 
$\hat Y^t_\nu$ to $\Om_{\nu,E}\times \U(k_\nu)$: its ergodicity with respect 
to the product measure that consists of Liouville measure on $\Om_{\nu,E}$ 
and Haar measure on $\U(k_\nu)$ implies quantum ergodicity. Since, however, 
the dynamics in the eigenspaces is completely fixed by a restriction to the
group $G$, the dynamical behaviour of the flow $\hat Y^t_\nu$ is determined
by that of $\tilde Y^t_\nu$. One hence concludes that in order 
to proof quantum ergodicity one requires the following condition (see 
Remark~\ref{rem:QEwithG}):
\begin{itemize}
\label{Irr} 
 \item[(Irr$_\nu$)] The representation $\rho: G \to \U(k_\nu)$ is irreducible.
\end{itemize}
In the sequel we always assume this to be the case.

Our intention now is to relate the dynamics in the eigenspaces, 
given by the conjugation with the transport matrices $d_{\nu\nu}$, to 
proper classical dynamics. To this end we require a symplectic manifold 
with the dynamics realised in a Hamiltonian fashion. For this purpose
Kirillov's orbit method \cite{Kir76} provides the necessary tools:
it relates the unitary irreducible representation $(\rho,\kz^{k_\nu})$ to a 
coadjoint orbit $\cO$ of $G$, which is a symplectic manifold. Moreover, the 
conjugation dynamics is realised in terms of the coadjoint action of $G$
on $\cO$. As in the case of $G=\SU(2)$ considered in \cite{BolGlaKep01},
this setting then also allows to introduce a Moyal-type quantisation such 
that hermitian matrix valued symbols can be uniquely related to real valued
functions on the symplectic product phase space $\TRd\times\cO$.

Let us now recall some properties of coadjoint orbits \cite{Kir76}:
The adjoint representation $\Ad: G \to \aut(\mathfrak{g})$, 
$g \mapsto (\ts_e \inn)(g)$, of a Lie group $G$ on its Lie algebra 
$\mathfrak{g}\cong \ts_e G$ is defined as the differential $\ts_e \inn$ 
of the inner automorphism $\inn(g): G \to G$, $x \mapsto g x g^{-1}$, 
$g \in G$, at the identity $e \in G$. The coadjoint representation of $G$ 
on the dual Lie algebra $\mathfrak{g}^\ast$ is then provided by the dual 
$\Ad^\ast_g:=(\Ad_{g^{-1}})^\ast$ of the linear map $\Ad_{g^{-1}}$, i.e.,
\begin{equation*}
 (\Ad^\ast_{g}(\lambda), X ) = (\lambda , \Ad_{g^{-1}} X),
\end{equation*}
for $X \in \mathfrak{g}$ and $\lambda \in \mathfrak{g}^\ast$; here 
$(\ , \ ): \mathfrak g^\ast \times \mathfrak g \to \rz$ denotes the dual
pairing between the vector spaces $\mathfrak{g}$ and $\mathfrak g^\ast$.
A coadjoint orbit $\cO_\lambda$ through $\lambda \in \mathfrak g^\ast$
then is an orbit of this group action,
\begin{equation*}
 \cO_\lambda:= \{ \Ad^\ast_{g} (\lambda);\ g \in G \} \subset 
 \mathfrak g^\ast.
\end{equation*}
If $G$ is compact, $\cO_\lambda$ is a smooth embedded and compact submanifold
of $\mathfrak g^\ast$. One of the main features of coadjoint orbits is their 
symplectic structure \cite{Kir76}.
\begin{prop} 
\label{thm:CoadjointOrbit}
Let $G$ be a connected Lie group and $\cO \subset \mathfrak g^\ast$ a 
coadjoint orbit. Then $\cO$ is a symplectic manifold and there exist unique 
symplectic forms $\sigma^\pm$ on $\cO$ such that
\begin{equation*}
 \si^\pm(\la) \bigl(\ad^\ast_X \la, \ad^\ast_Y \la \bigr) =  
 \pm \bigl(\la, [X,Y]\bigr)
\end{equation*}
for all $\la \in \cO$ and $X,Y \in \mathfrak g$. Here $\ad^\ast$ denotes
the differential of the coadjoint action and $[\ , \ ]$ is the Lie bracket 
on $\mathfrak g$. The forms $\sigma^\pm$ are referred to as the coadjoint 
orbit symplectic structures.
\end{prop}
Furthermore, let $G_\la:= \{g \in G;\ \Ad^\ast_{g} \la = \la \}$ denote the 
isotropy subgroup of $\la \in \mathfrak g^\ast$ under the coadjoint action.
Then this is a closed subgroup of $G$, and so the quotient $G/G_\lambda$ 
is a smooth manifold with smooth projection $\pi: G \to G/G_\la$ such that
one can identify $G/G_\lambda \cong \cO_\lambda$ via the diffeomorphism 
$\kappa:g G_\lambda \mapsto \Ad^\ast_{g} \lambda$. Moreover, since the
coadjoint action on $\cO_\la$ preserves its symplectic structure and is 
obviously transitive, $\cO_\la$ is a symplectic homogeneous space. In the 
following we only need one symplectic structure that turns $\cO_\la$ into a 
symplectic homogeneous space and therefore now fix $\si :=\si^+$.  

Our next goal is to construct a certain quantisation of the symplectic 
manifold $\cO_\la$. On the classical side one considers suitable 
functions on the phase
space $\cO_\la$ as observables; here we choose functions that are integrable
with respect to the volume form $\ud\eta$ that arises as the maximal exterior
power of the symplectic form $\si$. A Hamiltonian dynamics is then generated 
by a smooth real valued function $h$ on $\cO_\la$ through the association of 
a Hamiltonian vector field $\cX_h$ according to $\si (\cX_h,\cdot) =\ud h$. 
On the quantum side observables are hermitian endomorphisms of the 
representation space $V$ with inner product $\langle\cdot,\cdot\rangle_V$. 
A Moyal quantiser now assigns to a hermitian $A\in\cL (V)$ a function 
$a\in L^1 (\cO_\la)$ such that $\rho(g)A\rho(g^{-1})$ is mapped to 
$a\circ\Ad^\ast_{g^{-1}}$. This covariance property then ensures that the 
dynamics given by the conjugation with $D_\nu =\rho(g_\nu)$ is represented 
on the phase space $\cO_\la$ through the coadjoint action of $g_\nu$. 
Quantisations of this type were constructed by Simon \cite{Sim80}, who 
introduced suitable Berezin symbols representing $A\in\cL (V)$. However, 
here we will closely follow \cite{FigGraVar90}, where it is demonstrated 
that one can obtain a quantisation with an additional tracial property that 
turns out to be useful later. 

In the present context $G$ is a matrix Lie group, i.e., a closed subgroup
of $\mathrm{GL}_n(\kz)$, and its Lie algebra $\mathfrak g$ is a subalgebra
of $\mat_n(\kz)$ with the matrix commutator as Lie bracket. Thus there also
exists a non-degenerate symmetric bilinear form $B$ on $\mathfrak g$, given 
by $B(X,Y)=\mathrm{Re}\,\mtr(XY)$, that allows to identify $\mathfrak g$ 
with its dual $\mathfrak g^\ast$. Consider now the unitary irreducible 
representation $(\rho,V)$ of $G$ and fix a highest (real) weight 
$\la\in \mathfrak t^\ast$ corresponding to this representation, where
$\mathfrak t$ is the Lie algebra of a suitable maximal torus $T\subset G$.
Since one can identify $\mathfrak g$ and $\mathfrak g^\ast$ via the bilinear
form $B$, one can regard the highest weight as $\la\in\mathfrak g^\ast$.
Up to a phase, to this highest weight there corresponds a unique normalised
weight vector $w_\lambda \in V$. Now define the map $J:V \to \mathfrak g^\ast$
by
\begin{equation*}
 (J(v), X) := \langle v, \ud \rho(X) v\rangle_V,
\end{equation*}
such that in particular $J(w_\la)=\la$. This map is equivariant in the sense 
that $J(\rho(g)v) = \Ad^\ast_{g} J(v)$. Since the weight space of the maximal 
weight is one-dimensional one obtains
\begin{equation*}
 J^{-1} (\lambda) = \{ z w_\lambda; \ z \in \kz,\ |z|=1 \},
\end{equation*} 
and therefore 
\begin{equation*}
 J^{-1}( \cO_\lambda) = \{ \rho(g) w_\lambda; \ g \in G \}.
\end{equation*}
This setting now allows to associate to points $\eta\in\cO_\la$ vectors
$v_\eta\in V$ that are unique up to a phase. For this purpose one chooses a
measurable section $\eta \mapsto g_\eta$ in $G \to G/G_\la$ with 
$\Ad^\ast_{g_\eta}(\la)=\eta$ and $g_\la=e$, which is possible due to the 
fact that $\cO_\la \cong G/G_\la$ is an orbit of the coadjoint action. 
Then define for every $\eta \in \cO_\lambda$ a vector 
$v_\eta := \rho(g_\eta) w_\la$, which can also be viewed as a coherent 
state \cite{Per86}. The equivariance of the map $J$ now implies that 
$J(v_\eta)=\eta$, such that $v_\eta$ is unique up to a phase. This finally 
allows to define for every $A\in\cL(V)$ the unique covariant symbol
\begin{equation} 
\label{eq:CovSymb}
 Q_A(\eta) := \langle v_\eta,A v_\eta\rangle_V.
\end{equation}
In fact, $Q_A :\cO_\la \to \kz$ is continuous. We denote the space of
covariant symbols that are constructed according to the above scheme by
$\cS_\lambda:=\{ Q_A;\ A \in \cL(V) \}$ and recall from 
\cite{FigGraVar90}
\begin{lemma} 
\label{prop:CovSymbMap}
The map $\mathrm \cL(V) \rightarrow \cS_\lambda$ defined in
equation (\ref{eq:CovSymb}) is one-to-one.
\end{lemma}
Now consider a normalised vector $w\in V$ and the associated orthogonal 
projector $\Pi_w$ onto $\kz w \subset V$. Since, in the language of quantum 
mechanics, expectations of an observable $A\in\cL(V)$ in the state $w$
read $\langle w,A w \rangle_V = \mtr (A\Pi_w)$, one would like a Moyal
quantisation to represent $\mtr (AB)$ as
\begin{equation*} 
 \int_{\cO_\la} \overline{Q}_A (\eta) Q_B (\eta)\ \ud\eta.
\end{equation*}
This relation, however, does not hold. Considering $\cL (V)$ as a 
(finite-dimensional) Hilbert space with inner product $\mtr(A^\ast B)$,
we are hence looking for an isometry $\cL(V) \to L^2(\cO_\la)$. To this
end one notices that $\cS_\la$ being a finite dimensional subspace of
$L^2(\cO_\la)$, the Riesz representation theorem ensures for every 
$A\in\cL(V)$ that the linear form $L_A:\cS_\la \to \kz$ given by 
$L_A (Q_B):=\mtr(A^\ast B)$ can be represented in terms of a unique function 
$P_A \in \cS_\la$ such that  
\begin{equation*} 
 L_A(Q_B) = \mtr(A^\ast B) =
 \int_{\cO_\la} \overline{P}_A (\eta) Q_B (\eta)\ \ud\eta.
\end{equation*}
Since according to Lemma \ref{prop:CovSymbMap} the spaces $\cL(V)$ and 
$\cS_\lambda$ have the same (finite) dimension, the map 
$\cL(V) \ni A \mapsto P_A \in \cS_\lambda$ is a (linear) bijection that
can as well serve as a symbol map; $P_A$ is then called contravariant symbol.
We remark that in order to satisfy the natural condition $P_{\id_{V}}=1$, 
the volume form on $\cO_\la$ has to be normalised such that 
$\vol(\cO_\lambda)=\dim V$. Both the covariant and the contravariant symbol 
of $A\in\cL(V)$ fulfill the covariance condition
\begin{equation*} 
\begin{split} 
 Q_{\rho(g) A \rho(g^{-1})}(\eta)  &= Q_A(\Ad^\ast_{g^{-1}} \eta), \\
 P_{\rho(g) A \rho(g^{-1})}(\eta)  &= P_A(\Ad^\ast_{g^{-1}} \eta),
\end{split} 
\end{equation*}
for all $g \in G$ and all $\eta \in \cO_\lambda$. However, in order to
obtain the desired isometry from $\cL(V)$ into $L^2(\cO_\la)$, one is forced
to introduce a symbol map that in a certain sense lies in between $Q$ and $P$. 
 
In \cite{FigGraVar90} it is shown that the operator $K$ on $\cS_\la$
that maps $Q_A$ to $P_A$ is bijective and positive. It therefore allows
for a (positive) square-root $K^{1/2}$ which can be used to define a
symbol map with all desired properties:
\begin{defn} 
\label{defn:StratWeylSymbol}
For $A \in \cL(V)$ the Stratonovich-Weyl symbol $ \symb^{SW}[A]\in \cS_\la$
is given by
\begin{equation*} 
 \symb^{SW}[A]:= K^{1/2} Q_A = K^{-1/2} P_A.
\end{equation*}
\end{defn}
Summarising the above finally yields \cite{FigGraVar90}:
\begin{prop}
\label{prop:SWsymbol}
The symbol map $A \mapsto \symb^{SW}[A]$ has the following properties:
\begin{itemize}
 \item[(i)] It is a linear one-to-one map from $\cL(V)$ to $\cS_\la$,
 \item[(ii)] $\symb^{SW}[A^\ast]=\overline{\symb^{SW}[A]}$,
 \item[(iii)] $\symb^{SW}[\id_V] = 1$,
 \item[(iv)] $\symb^{SW}[\rho(g) A \rho(g^{-1})](\eta) = 
             \symb^{SW}[A](\Ad^\ast_{g^{-1}} \eta)$ for all 
             $\eta \in \cO_\lambda,\ g \in G$,
 \item[(v)] $\displaystyle \int_{\cO_\la} \symb^{SW}[A](\eta) 
            \symb^{SW}[B](\eta)\ {\rm d}\eta = \mtr(AB)$.
\end{itemize}
\end{prop}
In order to make the relation between $A\in\cL(V)$ and its symbol 
$\symb^{SW}[A]$ explicit, one introduces a (hermitian) quantiser 
$\De_\la : \cO_\la \to \cL(V)$ such that
\begin{equation}
\label{eq:quantiser}
\symb^{SW}[A] = \mtr \bigl( A\De_\la \bigr) \quad\text{and}\quad
A = \int_{\cO_\la} \symb^{SW}[A](\eta)\De_\la(\eta)\ \ud\eta.
\end{equation}
As shown in \cite{FigGraVar90}, the quantiser can be expressed in terms
of generalised spherical harmonics associated with those unitary irreducible
representations that appear in the regular representation 
$(\tau(g)f)(\eta)=f(\Ad^\ast_{g^{-1}}\eta)$ of $G$ on $\cS_\la$.

With this formalism at hand one can now transfer the dynamics of a (hermitian)
$B\in\cL(V)$ given by a conjugation with $D(t)=\rho(g(t))$, $B\mapsto B(t)
= D^{-1}(t)BD(t)$, to the coadjoint action of $g(t)$ on the symplectic manifold
$\cO_{\lambda}$ via the relation $\symb^{SW}[B(t)](\eta) =
\symb^{SW}[B](\Ad^\ast_{g(t)} \eta)$. The symplectic structure on $\cO_\la$ 
defined by the form $\si$, furthermore, allows to identify the dynamics 
$\eta \mapsto \Ad^\ast_{g(t)} \eta$ as being Hamiltonian. To see this
assume that $D(t)$ is determined by 
\begin{equation}
\label{eq:Ddyn}
\dot{D}(t) +\ui H D(t) =0 \quad\text{with}\quad D(0)=\id_V,
\end{equation}
where $H\in\cL(V)$ is hermitian; compare equation (\ref{eq:DDyn}). On
the one hand now, a Hamiltonian flow $\eta \mapsto \eta(t)$ can be introduced
on $\cO_\la$ that is generated by the Stratonovich-Weyl symbol of $H$. The 
associated Hamiltonian vector field $\cX_{\symb^{SW}[H]}$ is then defined 
through 
\begin{equation*}
\si(\cX_{\symb^{SW}[H]},\cdot)=\ud\symb^{SW}[H], 
\end{equation*}
so that the time evolution $f(t)(\eta)=f(\eta(t))$ of a function 
$f\in C^\infty(\cO_\la)$ is governed by the equation
\begin{equation}
\label{eq:Difff(t)}
\dot{f}(t) = \bigl\{ \symb^{SW}[H],f(t) \bigr\} = 
\si \bigl( \cX_{\symb^{SW}[H]},\cX_{f(t)} \bigr) = 
-\ud f(t)\bigl( \cX_{\symb^{SW}[H]} \bigr).
\end{equation}
On the other hand, differentiating $\Ad^\ast_{g(t)} \eta$ with respect to
$t$ yields
\begin{equation}
\label{eq:Diffcoadj}
\frac{\ud}{\ud t}\left. \Ad^\ast_{g(t)} \eta\right|_{t=0} = -\ad^*_{X_H}\eta,
\end{equation}
where $X_H\in{\mathfrak g}$ is the generator of the curve $g(t)$ in G which, 
according to equation (\ref{eq:Ddyn}), is related to $H$ via 
$\ud\rho(X_H)=-\ui H$. A comparison of (\ref{eq:Difff(t)}) and 
(\ref{eq:Diffcoadj}) then shows that the dynamics provided by the coadjoint 
action $\Ad^\ast_{g(t)}$ coincides with the Hamiltonian flow generated by 
the symbol $\symb^{SW}[H]$.

As an ultimate outcome of the above formalism we are now in a position to
introduce a skew-product flow on the symplectic phase space
$\TRd \times \cO_\la$ that completely determines the time evolution of the
$\nu$-th diagonal block of an observable on the level of its principal
symbol. Explicitly, this flow is given by
\begin{equation}
\label{eq:skewO}
Y_\nu ^t : \TRd \times \cO_\la \to \TRd \times \cO_\la
\end{equation}
with 
\begin{equation}
\label{eq:skewflowO}
Y_\nu ^t (x,\xi,\eta) := \bigl( \Phi^t_\nu(x,\xi),\Ad^\ast_{g_\nu(x,\xi,t)}
\eta \bigr);
\end{equation}
it leaves the product measure $\ud x\,\ud\xi\,\ud\eta$ invariant.

Consider now a semiclassical pseudodifferential operator $\cB$ with symbol 
$B\in\scl^\infty(1)$. Mod $O(\hbar^\infty)$ the quantum dynamics preserves the 
diagonal structure of its blocks $\cP_\nu\cB\cP_\nu$. According to the Egorov 
theorem \ref{thm:Egorov}, together with the definition (\ref{eq:DefD}), 
the principal symbol of $\cP_\nu\cB(t)\cP_\nu$ hence reads
\begin{equation}
\label{eq:EgorovV}
V_\nu (x,\xi)D_\nu^\ast (x,\xi,t) \bigl( V_\nu^\ast B_0 V_\nu \bigr)
\bigl(\Phi^t_\nu(x,\xi)\bigr) D_\nu (x,\xi,t)V_\nu^\ast (x,\xi).
\end{equation}
We now exploit the possibility, explicitly provided by (\ref{eq:quantiser}), 
to uniquely represent the value of 
$V_\nu^\ast B_0 V_\nu :\TRd \to \cL(\kz^{k_\nu})$ in terms of a 
Stratonovich-Weyl symbol,
\begin{equation}
\label{eq:SWsymbdef}
b_{0,\nu} (x,\xi,\eta) := \symb^{SW}\bigl[ (V_\nu^\ast B_0 V_\nu)(x,\xi) 
\bigr] (\eta).
\end{equation}
The dynamics of the principal symbol in this representation is now 
summarised in the following variant of the Egorov theorem:
\begin{prop} \label{prop:SWEgorov}
The Stratonovich-Weyl symbol $b(t)_{0,\nu}$ associated with the principal 
symbol of the operator $\cP_\nu\cB(t)\cP_\nu$ is the time evolution of
$b_{0,\nu}$ under the skew-product flow $Y^t_\nu$ defined in equations 
(\ref{eq:skewO})--(\ref{eq:skewflowO}), i.e.,
\begin{equation*}
b(t)_{0,\nu} (x,\xi,\eta) = b_{0,\nu} \bigl( Y^t_\nu(x,\xi,\eta) \bigr).
\end{equation*}
\end{prop}
\begin{proof}
According to (\ref{eq:EgorovV}) and (\ref{eq:SWsymbdef}), $b(t)_{0,\nu}$ 
is given by 
\begin{equation*}
b(t)_{0,\nu} (x,\xi,\eta) = \symb^{SW} \bigl[ \rho(g_\nu^{-1}(x,\xi,t)) 
\bigl( V_\nu^\ast B_0 V_\nu \bigr)\bigl(\Phi^t_\nu(x,\xi)\bigr) 
\rho (g_\nu (x,\xi,t)) \bigr](\eta),
\end{equation*}
which due to the covariance property $(iv)$ of Proposition 
\ref{prop:SWsymbol} reads
\begin{equation*}
\begin{split}
b(t)_{0,\nu} (x,\xi,\eta) 
 &= \symb^{SW} \bigl[ \bigl( V_\nu^\ast B_0 V_\nu \bigr)\bigl(
    \Phi^t_\nu(x,\xi)\bigr) \bigr](\Ad^\ast_{g_\nu (x,\xi,t)}\eta)  \\
 &=b(t)_{0,\nu} \bigl(\Phi^t_\nu(x,\xi),\Ad^\ast_{g_\nu(x,\xi,t)}\eta\bigr).
\end{split}
\end{equation*}

\end{proof}

%% file: 5sec.tex
\section{Trace asymptotics and a limit formula for averaged 
expectation values} 
\label{sec:TraceAsymptotics}
A fundamental ingredient in the asymptotics of eigenvectors we are
aiming at is a semiclassical limit formula for the expectation values of 
bounded operators $\cB$ on $L^2(\rz^d) \otimes \kz^n$. Below we will obtain 
a Szeg\"o-type formula which connects semiclassically averaged expectation 
values with objects that can be calculated from the principal symbol $B_0$ 
of the operator $\cB$ and therefore allow for a classical interpretation.
On the so defined classical side we fix a value $E$ for all eigenvalue
functions $\la_\nu$, $\nu=1,\dots,l$, of the principal symbol $H_0$ with
the following properties:
\begin{itemize}
\label{H3-4}
\item[(H3$_\nu$)] 
 There exists some $\ve>0$ such that all 
 $\la_{\nu}^{-1}([E-\ve ,E+\ve])\subset\TRd$ are compact.
\item[(H4$_\nu$)] The functions $\lambda_\nu$ shall possess no critical 
 values in $[E-\varepsilon,E+\varepsilon]$.
\item[(H5$_\nu$)] Among the level surfaces $\Om_{\nu,E}=\la_\nu^{-1}(E)$,
 $\nu=1,\dots,l$, at least one is non-empty.
\end{itemize}
In addition to (H1) and (H2), which imply the essential selfadjointness of
the operator $\cH$, these conditions guarantee as in the scalar case 
\cite{DimSjo99} that for sufficiently 
small $\hbar$ the spectrum of $\cH$ is discrete in any open interval
contained in $[E-\ve ,E+\ve]$. This setting now allows us to generalise 
the constructions made in \cite{BolGla00} to Hamiltonians with non-scalar 
principal symbols: The expectation values of an operator $\cB$ will be 
considered in normalised eigenvectors $\psi_j$ of $\cH$ with corresponding 
eigenvalues $E_j$ in an interval $I(E,\hbar)=[E-\hbar \omega,E+\hbar \omega]$,
$\omega>0$, such that $I(E,\hbar) \subset [E-\ve,E+\ve]$ if $\hbar$ is small 
enough. On the classical side the Hamiltonian flows $\Phi^t_\nu$ generated by
the eigenvalue functions $\la_\nu$ will enter on the level surfaces 
$\Om_{\nu,E}$. Regarding these we assume:
\begin{itemize}
 \item[(H6$_\nu$)] The periodic points of $\Phi^t_\nu$ with non-trivial 
   periods form a set of Liouville measure zero in $\Omega_{\nu,E}$.
\end{itemize}
The quantities appearing on the classical side of the limit formula 
turn out to be averages of smooth matrix valued functions 
$B \in C^\infty(\TRd)\otimes\mat_n(\kz)$ over $\Om_{\nu,E}$ with respect to 
Liouville measure, for which we introduce the notation
\begin{equation*}
 \ell_{\nu,E}(B):= \int_{\Omega_{\nu,E}} B(x,\xi)\ \dl(x,\xi).
\end{equation*}
The main result of this section is now summarised in the following 
Szeg\"o-type limit formula:
\begin{prop} 
\label{prop:Szego}
Let $\cH$ be a semiclassical pseudodifferential operator with symbol
$H\in\scl^0(m)$, such that the principal symbol $H_0$ satisfies the 
assumptions (H0)--(H2) and (H3$_\nu$)--(H6$_\nu$) for all $\nu=1,\ldots,l$. 
Furthermore, let $\cB$ be an operator with symbol $B \in \scl^0(1)$ and 
principal symbol $B_0$. Then the limit formula
\begin{equation} 
\label{eq:Szego}
 \lim_{\hbar \to 0} \frac{1}{N_I} \sum_{E_j \in I(E,\hbar)}
 \langle \psi_j, \cB \psi_j \rangle = \frac{\sum_{\nu=1}^l \vol \Omega_{\nu,E}
 \mtr \ell_{\nu,E}(P_{\nu,0} B_0 P_{\nu,0})}
 {\sum_{\nu=1}^l k_\nu \vol \Omega_{\nu,E}}
\end{equation}
holds.
\end{prop}
\begin{proof}
Adapted to the spectral localisation mentioned above we choose a smooth and 
compactly supported function $g \in C_0^\infty(\rz)$ such that $g(\la)=\la$ 
on a neighbourhood of $[E-\ve,E+\ve]$. Furthermore, we apply the semiclassical
splitting of the Hilbert space $L^2(\rz^d) \otimes \kz^n$ given by the 
projection operators $\cP_\nu$,
\begin{equation}
\label{eq:L2decomp}
 L^2(\rz^d) \otimes \kz^n = \ran\cP_1 \oplus \cdots \oplus
 \ran\cP_l \quad \text{mod} \quad \hbar^\infty,
\end{equation} 
and the corresponding decomposition $\cH=\sum_{\nu=1}^l \cH \cP_{\nu}$ 
(mod $O(\hbar^\infty)$) of the Hamiltonian. By employing the generalisation 
of the Helffer-Sj\"ostrand formula to matrix valued operators developed in 
\cite{Dim93,Dim98}, we represent $g(\cH)=\sum_{\nu=1}^l g(\cH \cP_\nu)\cP_\nu$
(mod $O(\hbar^\infty)$) with 
\begin{equation*}
 g(\cH \cP_\nu) \cP_\nu = -\frac{1}{\pi} \int_{\kz} \dpr_{\overline{z}} 
 \tilde g(z) (\cH-z)^{-1} \cP_\nu \ \ud z,
\end{equation*}
where $\tilde g$ is an almost-analytic extension of $g$. Since the principal
symbol $H_0 P_{\nu,0}$ of $\cH \cP_\nu$ is scalar, 
$H_0 P_{\nu,0}=\la_\nu P_{\nu,0}$, when considered to act on sections in 
the eigenvector bundle $E^\nu$, one can use the methods of \cite{DimSjo99} 
to show that on $\la_\nu^{-1}([E-\ve,E+\ve])$ the asymptotic expansions of 
$\symb^W [g(\cH\cP_\nu)]$ and of $\symb^W [\cH \cP_\nu]$ coincide. Below we 
will always employ the spectral localisation to the interval $I(E,\hbar)$, and 
since $\symb^W [g(\cH \cP_\nu)] \in \scl^0(1)$, one can therefore now assume 
that $H \in \scl^0(1)$. Furthermore, the decomposition (\ref{eq:L2decomp}) 
allows us to employ the techniques of \cite{DimSjo99} in the same manner as 
in \cite{BolGla00}. Hence, if $\chi \in C_0^\infty(\rz)$ with $\chi \equiv 1$ 
on $I(E,\hbar)$ and $\supp \chi \subset [E-\varepsilon,E+\varepsilon]$, the 
operator
\begin{equation*} 
 \cU_\chi(t):= \ue^{- \frac{\ui}{\hbar} \cH t} \chi(\cH) \sum_{\nu=1}^l 
 \cP_\nu = \sum_{\nu=1}^l \ue^{- \frac{\ui}{\hbar} \cH \cP_\nu t} 
 \chi( \cH\cP_\nu) \cP_\nu \quad \text{mod} \quad O(\hbar^\infty),
\end{equation*} 
has a pure point spectrum. Moreover, each of the operators 
$\ue^{-\frac{\ui}{\hbar}\cH\cP_{\nu}t}\chi(\cH\cP_\nu)$ can be approximated 
in trace norm up to an error of $O(\hbar^\infty)$ by a semiclassical Fourier 
integral operator with a kernel of the form
\begin{equation} 
\label{eq:SemApproxU}
 K_\nu (x,y,t) = 
 \frac{1}{(2 \pi \hbar)^d} \int_{\rz^d} a_\nu(x,y,t,\xi) 
 \ue^{\frac{\ui}{\hbar}(S_\nu(x,\xi,t)- \xi y )}\ \ud \xi.
\end{equation} 
Here, as in \cite{BolKep99a}, the phases $S_\nu$ have to fulfill the 
Hamilton-Jacobi equations
\begin{equation*} 
 \lambda_\nu\bigl(x,\dpr_x S_\nu(x,\xi,t)\bigr) + \dpr_t S_\nu(x,\xi,t)=0, 
 \quad S_\nu(x,\xi,0)=x \xi.
\end{equation*}
The amplitudes $a_\nu \in \scl^0(1)$ with asymptotic expansions 
$a_\nu \sim \sum_{j=0}^\infty \hbar^j a_{\nu,j}$ are determined as solutions 
of certain transport equations \cite{BolKep99a} with initial conditions
$\left. a_\nu \right|_{t=0}=\chi(\lambda_\nu) P_{\nu,0} + O(\hbar)$. 
Following \cite{BolGla00} further, we choose test functions 
$\rho \in C^\infty(\rz)$ with compactly supported Fourier transforms 
$\hat\rho \in C_0^\infty(\rz)$ such that 
\begin{equation*}
 \tr \frac{1}{2 \pi} \int_\rz \hat \rho(t) \ue^{\frac{\ui}{\hbar}  
 E t} \cB \cU_\chi(t)\ \ud t = \sum_{j} \chi(E_j) 
 \langle \psi_j, \cB \psi_j \rangle \rho \left( \frac{E_j-E}{\hbar} \right),
\end{equation*} 
where $\tr$ denotes the operator trace on the Hilbert space $L^2(\rz^d)
\otimes \kz^n$. Using the semiclassical approximation (\ref{eq:SemApproxU})
one now has to calculate 
\begin{equation} 
\label{eq:LeadingOrderSzego}
 \frac{1}{2 \pi (2 \pi \hbar)^d} \int_\rz \int_{\rz^d} \int_{\rz^d} 
  \hat \rho(t) \sum_{\nu=1}^l \mtr \bigl( B_0(x,\dpr_x S_\nu)
  a_{\nu,0}(x,x,t,\xi) \bigr) \ue^{\frac{\ui}{\hbar} (S_\nu(x,\xi,t) - x \xi 
 +E t)}\ \ud \xi\,\ud x\,\ud t
\end{equation} 
in leading semiclassical order. This can be done with the method of stationary
phase, where the stationary points 
$(x_{\nu,\text{st}},\xi_{\nu,\text{st}},t_{\nu,\text{st}})$ of the phase 
$S_\nu(x,\xi,t)-x \xi + Et$ determine periodic points 
$(x_{\nu,\text{st}},\xi_{\nu,\text{st}}) \in \Om_{\nu,E}$ of the Hamiltonian 
flow $\Phi^t_\nu$ with periods $t_{\nu,\text{st}}$. Since the eigenvalue 
function $\lambda_\nu$ is supposed to be non-critical at $E$, the periods 
$t_{\nu,\text{st}}$ of the flow $\Phi^t_\nu$ cannot accumulate at zero, see 
\cite{Rob87}. One can hence split $\hat\rho = \hat\rho_1 + \hat\rho_2$ 
in such a way that $\hat \rho_1$ is supported only in a small neighbourhood 
of zero and $\hat\rho_2 =0$ in the vicinity of zero, so that the only 
period in $\supp \hat \rho_1$ is the trivial one, $t_{\nu,\text{st}}=0$.  
The contribution coming from $\hat\rho_1$ to (\ref{eq:LeadingOrderSzego}) 
is therefore determined by the periodic points with $t_{\nu,\text{st}}=0$. 
These build up the entire level surface $\Om_{\nu,E}$ which, according to 
assumption (H3$_\nu$), is compact. The result then reads 
(see \cite{DimSjo99,BolGla00})
\begin{equation} 
\label{eq:LeadingOrderSzego1}
 \sum_j \chi(E_j) \langle \psi_j,\cB \psi_j \rangle \rho_1 \left( 
 \frac{E_j-E}{\hbar} \right) = \chi(E) \frac{\hat \rho_1(0)}{2 \pi} 
 \sum_{\nu=1}^l \frac{\vol \Omega_{\nu,E}}{(2 \pi \hbar)^{d-1}} 
 \bigl( \mtr \ell_{\nu,E}( P_{\nu,0} B_0 P_{\nu,0}) + O(\hbar) \bigr).
\end{equation}
Coming to the contribution of the term with $\hat\rho_2$ to the expression
(\ref{eq:LeadingOrderSzego}), we recall that $\hat \rho_2$ has been chosen 
to vanish in a neighbourhood of zero. The relevant stationary points are 
hence related to periodic orbits of the flow $\Phi^t_\nu$ with non-vanishing 
periods. The condition (H6$_\nu$) now allows us to employ the methods of 
\cite{DimSjo99}, leading to the estimate
\begin{equation} 
\label{eq:LeadingOrderSzego2}
 \sum_{j} \chi(E_j) \langle \psi_j,\cB \psi_j \rangle \rho_2 
 \left( \frac{E_j-E}{\hbar} \right) = o(\hbar^{1-d}).
\end{equation}
The relations (\ref{eq:LeadingOrderSzego1}) and (\ref{eq:LeadingOrderSzego2})
together therefore imply that for every test function $\rho\in C^\infty(\rz)$
with Fourier transform $\hat\rho\in C_0^\infty(\rz)$ the estimate 
(\ref{eq:LeadingOrderSzego1}) holds with $\rho_1$ replaced by $\rho$.
Hence, the Tauberian argument developed in \cite{BruPauUri95} can be 
applied to yield
\begin{equation} \label{eq:LeadingOrderSzego1&2}
 \sum_{E_j \in I(E,\hbar)} \langle \psi_j, \cB \psi_j \rangle 
 = \frac{\omega}{\pi} \sum_{\nu=1}^l \frac{\vol \Omega_{\nu,E}}{
 (2 \pi \hbar)^{d-1}} \mtr \ell_{\nu,E}(P_{\nu,0} B_0 P_{\nu,0})
 + o(\hbar^{1-d}).
\end{equation}
In this relation one can set the operator $\cB$ equal to the identity and 
thus obtains a semiclassical expression for the number $N_I$ of eigenvalues 
of $\cH$ in $I(E,\hbar)$,
\begin{equation} 
\label{eq:N_I}
 N_I := \# \{ E_j\in I(E,\hbar) \} = \frac{\omega}{\pi} \sum_{\nu=1}^l 
 k_\nu \frac{\vol \Omega_{\nu,E}}{(2 \pi \hbar)^{d-1}} + o(\hbar^{1-d}),
\end{equation}
where $k_\nu=\mtr P_{\nu,0}$ denotes the dimension of the fibre 
$\ran P_{\nu,0}=E^\nu$ corresponding to the eigenvalue $\la_\nu$ of $H_0$.
The proof is now finished by combining the expressions 
(\ref{eq:LeadingOrderSzego1&2}) and (\ref{eq:N_I}).
\end{proof}
\noindent Let us add two comments:
\begin{enumerate}
\item Under the additional assumption (Irr$_\nu$) the 
Stratonovich-Weyl calculus discussed in section \ref{sec:eigendyn} can be
applied. It allows to express $\mtr (P_{\nu,0} B_0 P_{\nu,0}) =
\mtr (V_\nu^\ast B_0 V_\nu)$ in terms of the symbol $b_{0,\nu}$ introduced in 
(\ref{eq:SWsymbdef}). This then leads to the representation
\begin{equation}
\label{eq:Mmeandef}
\begin{split}
 \frac{1}{k_\nu} \mtr \ell_{\nu,E}( P_{\nu,0} B_0 P_{\nu,0}) 
 &= \frac{1}{\vol\cO_\la} \int_{\Om_{\nu,E}}\int_{\cO_\la}b_{0,\nu}
    (x,\xi,\eta)\ \ud\eta\,\dl(x,\xi) \\ 
 &=: \text{M}_{E,\nu,\la} \bigl( b_{0,\nu} \bigr)
\end{split}
\end{equation}
as an integral over the product space $\Om_{\nu,E}\times\cO_\la$. Here the 
relation $k_\nu = \vol\cO_\la$, introduced in section~\ref{sec:eigendyn}, 
enables one to give the right-hand side of (\ref{eq:Szego}) a genuinely 
classical interpretation.
\item The operators $\cB$ considered in the limit formula (\ref{eq:Szego}) 
have not been restricted to those with symbols in the invariant subalgebra 
$\sinv^0(1)\subset\scl^0(1)$. Nevertheless, only the diagonal blocks of 
their principal symbols $B_0$ with respect to the projection matrices 
$P_{\nu,0}$ enter on the right-hand side of (\ref{eq:Szego}). In particular, 
this implies that for an operator $\cB$ with a purely off-diagonal 
principal symbol, i.e., $P_{\mu,0} B_0 P_{\mu,0}=0$ for all $\mu=1,\ldots,l$, 
the semiclassical average vanishes. Thus one can replace an operator $\cB$ 
with symbol $B\in\scl^0(1)$ by its diagonal part $\sum_\mu \tilde\cP_\mu
\cB \tilde\cP_\mu$, whose symbol is in the invariant algebra $\sinv^0(1)$, 
without changing the value of the limit on the right-hand side of 
(\ref{eq:Szego}).
\end{enumerate}
So far we have considered expectation values in normalised eigenvectors of 
$\cH$. Our intention now is to discuss the projections $\cP_\nu \psi_j$ 
of the eigenvectors of $\cH$ to a fixed almost invariant subspace of 
$L^2(\rz^d)\otimes\kz^n$. One thus expresses averaged expectation values in
the projected eigenvectors in terms of classical quantities related to
the single Hamiltonian flow $\Phi^t_\nu$. In order to achieve this one
applies Proposition \ref{prop:Szego} to operators $\cP_\nu\cB\cP_\nu$
and exploits the selfadjointness of $\cP_\nu$. This results in 
\begin{cor}
\label{cor:restrictSzegoe}
Under the assumptions stated in Proposition \ref{prop:Szego}, for each
$\nu\in\{1,\dots,l\}$ the restricted limit formula
\begin{equation}
\label{eq:Szegoenu}
 \lim_{\hbar \to 0} \frac{1}{N_I} \sum_{E_j \in I(E,\hbar)} 
 \langle \cP_\nu \psi_j, \cB \cP_\nu \psi_j \rangle = 
 \frac{ \vol \Omega_{\nu,E} \mtr \ell_{\nu,E} (P_{\nu,0} B_0 
 P_{\nu,0})}{\sum_{\mu=1}^l k_\mu \vol \Omega_{\mu,E}}.
\end{equation}
holds.
\end{cor}
Thus the semiclassical average of the projected eigenvectors 
$\cP_\nu\psi_j$, with $E_j\in I(E,\hbar)$, localises on the 
corresponding level surface $\Om_{\nu,E}\subset\TRd$. If one considers
(\ref{eq:Szegoenu}) for different $\nu$, the relative weights 
of the corresponding projections are determined by the relative volumes of 
the associated level surfaces and the dimensions of the eigenspaces $E^\nu$,
which equal the volumes of the coadjoint orbits $\cO_\la$. 

In general, however, the projected eigenvectors $\cP_\nu\psi_j$ are 
neither normalised, nor are they genuine eigenvectors of $\cH$. We therefore
now introduce the normalised vectors 
\begin{equation}
\label{eq:Defquasim}
\phi_{j,\nu} := \frac{\cP_\nu\psi_j}{\|\cP_\nu\psi_j\|}.
\end{equation}
Since the projectors $\cP_\nu$ only commute with $\cH$ up to a term of 
$O(\hbar^\infty)$, the pairs $(E_j,\phi_{j,\nu})$ are quasimodes with 
discrepancies $r_{j,\nu}$, i.e.,
\begin{equation*}
\bigr( \cH - E_j \bigl) \phi_{j,\nu} = 
\frac{[\cH,\cP_\nu]\psi_j}{\|\cP_\nu\psi_j\|} 
\quad\text{and}\quad 
r_{j,\nu} = \frac{\| [\cH,\cP_\nu]\psi_j\|}{\|\cP_\nu\psi_j\|}.
\end{equation*}
This observation only ensures the existence of an eigenvalue of $\cH$
in the interval $[E_j-r_{j,\nu},E_j+r_{j,\nu}]$, which is a trivial
statement; it does not imply that $\phi_{j,\nu}$ is close to an
eigenvector of $\cH$, see \cite{Laz93}. It therefore is of somewhat more 
interest to consider the operator $\cH\cP_\nu$, whose spectrum 
inside the interval $[E-\ve,E+\ve]\supset I(E,\hbar)$ is as well purely 
discrete. Following the above reasoning, one then concludes that 
$(E_j,\phi_{j,\nu})$ is a quasimode with discrepancy $r_{j,\nu}$ also for 
this operator. Thus, if $\|\cP_\nu\psi_j\|\ge c\hbar^N$ for some $N\geq 0$
and hence $r_{j,\nu}=O(\hbar^\infty)$, the operator $\cH\cP_\nu$ has an 
eigenvalue with distance $O(\hbar^\infty)$ away from $E_j$. Since there
are $N_I$ eigenvalues $E_j\in I(E,\hbar)$ one finds as many quasimodes for
$\cH\cP_\nu$. But this operator has only
\begin{equation*}
N_I^\nu = \frac{k_\nu\omega}{\pi}\frac{\vol\Om_{\nu,E}}{(2 \pi \hbar)^{d-1}} 
+ o(\hbar^{1-d})
\end{equation*}
eigenvalues in $I(E,\hbar)$, compare (\ref{eq:N_I}). This observation might
suggest that only approximately $N_I^\nu$ of the $N_I$ projected eigenvectors
$\cP_\nu\psi_j$ are of considerable size, such that the discrepancies
of the associated quasimodes are smaller than the distance of $E_j$ to
neighbouring eigenvalues of $\cH$. This expectation can be strengthened by 
an application of the limit formula (\ref{eq:Szegoenu}) with the choice 
$\cB=\id$,
\begin{equation}
\label{eq:Szegoeproject}
\lim_{\hbar\to 0} \frac{1}{N_I}\sum_{E_j\in I(E,\hbar)}
\|\cP_\nu\psi_j\|^2 = \frac{k_\nu\vol\Om_{\nu,E}}
{\sum_{\mu=1}^l k_\mu\vol\Om_{\mu,E}},
\end{equation}
which implies that
\begin{equation}
\label{eq:NInulimit}
N_I^\nu = \sum_{E_j\in I(E,\hbar)}\|\cP_\nu\psi_j\|^2 + o(1),
\quad \hbar\to 0.
\end{equation}
One could thus expect that roughly $N_I^\nu$ of the projected eigenvectors 
$\cP_\nu\psi_j$ are close to $\psi_j$, and the rest is such that 
$\|\cP_\nu\psi_j\|$ is semiclassically small. However, (\ref{eq:NInulimit}) 
does not rule out the other extreme situation, provided by projected 
eigenvectors $\cP_\nu\psi_j$, $\nu=1,\dots,l$, equidistributing in the sense 
that their squared norms are asymptotic to $N^\nu_I/N_I$ as $\hbar\to 0$. In 
that case the discrepancies of the associated quasimodes for the operators
$\cH\cP_\nu$ can be estimated as $r_{j,\nu}=O(\hbar^\infty)$. In order 
now that these quasimodes do not produce more than $N^\nu_I$ eigenvalues
of $\cH\cP_\nu$ in $I(E,\hbar)$, a finite fraction of the eigenvalues
$E_j$ of $\cH$ must possess spacings to their nearest neighbours of the
order $\hbar^\infty$. Since in general there exist no sufficient lower 
bounds on eigenvalue spacings, none of the two extreme situations discussed
above can be excluded so far.

What is possible, however, is to derive from (\ref{eq:Szegoeproject}) an
upper bound for the fraction of the projected eigenvectors $\cP_\nu\psi_j$ 
that are close in norm to $\psi_j$,
\begin{equation*} 
\lim_{\hbar\to 0}\frac{1}{N_I} \# \bigl\{ E_j\in I(E,\hbar);\ \| \cP_\nu
\psi_j -\psi_j \| = o(1) \bigr\} \leq 
\frac{k_\nu\vol\Om_{\nu,E}}{\sum_{\mu=1}^l k_\mu\vol\Om_{\mu,E}},
\end{equation*} 
see also \cite{Sch01}.
To obtain lower bounds is notoriously more difficult. The limit formula
(\ref{eq:Szegoeproject}) only allows to estimate the fraction of projected
eigenvectors with norms that tend to a finite limit as $\hbar\to 0$. One
conveniently measures this fraction in units of the value that is
expected for equidistributed projections. Therefore, with $\de := \tilde\de
\frac{k_\nu \vol\Om_{\nu,E}}{\sum_{\mu=1}^l k_\mu \vol\Om_{\mu,E}}$,
we consider
\begin{equation*}
 N^\delta_{\nu,I}:= \# \bigr\{ E_j \in I(E,\hbar); \ \| \cP_\nu \psi_j \|^2
 \ge \de \bigl\}.
\end{equation*}
Since
\begin{equation*} 
\begin{split} 
 \frac{1}{N_I} \sum_{E_j \in I(E,\hbar)} \| \cP_\nu \psi_j \|^2 
 & \le \frac{1}{N_I} \sum_{\substack{E_j \in I(E,\hbar) \\ 
   \|\cP_{\nu} \psi_j\|^2 \ge \de}} 1 +  \frac{1}{N_I} 
   \sum_{\substack{E_j\in I(E,\hbar) \\ \|\cP_\nu \psi_j \|^2 < \de}} 
   \| \cP_\nu \psi_j \|^2 \\
 & \le  \frac{N_{\nu,I}^\de}{N_I} + \frac{\de}{N_I}(N_I -N_{\nu,I}^\de),
\end{split} 
\end{equation*}
the relative fraction of projected eigenvectors with finite semiclassical
limit can be estimated from below as
\begin{equation} 
\label{eq:Nde/N}
 \lim_{\hbar\to 0} \frac{N^\de_{\nu,I}}{N_I} \ge \frac{(1-\tilde\de) k_\nu 
 \vol \Om_{\nu,E}}{\sum_{\mu=1}^l k_\mu \vol\Om_{\mu,E}}.
\end{equation}

%% file: 6sec.tex
\section{Quantum ergodicity} 
\label{sec:EquiDist}
Our intention in this section is to consider quantum ergodicity for the 
normalised eigenvectors $\psi_j$, $E_j\in I(E,\hbar)$, of the quantum 
Hamiltonian $\cH$. In the case of scalar pseudodifferential operators one 
denotes by quantum ergodicity a weak convergence of the phase space lifts 
of almost all eigenfunctions to Liouville measure on the level surface 
$\Om_E=H_0^{-1}(E)$, and proves this to hold if the flow generated by the 
principal symbol $H_0$ of the quantum Hamiltonian is ergodic on $\Om_E$. 
In the present situation of operators with matrix valued symbols, however, 
each eigenvalue $\la_\nu$ of $H_0$ defines its own classical dynamics. One 
hence can only expect quantum ergodicity to be concerned with statements 
about the projections $\cP_\nu\psi_j$ of the eigenvectors to the different 
almost invariant subspaces of $L^2(\rz^d)\otimes\kz^n$ in relation to the 
behaviour of the associated classical systems. In the preceding section we 
discussed the question of identifying those projected eigenvectors whose norms 
are not semiclassically small. Since presently this problem cannot be resolved
directly, quantum ergodicity can only be formulated by restricting to
those eigenvectors whose squared norms exceed a value of $\de$ in the 
semiclassical limit, without specifying them further.

Conventionally the convergence of quantum states determined by the 
eigenvectors $\psi_j$ of $\cH$ is discussed in terms of expectation values
of observables in these states. Explicit lifts of the eigenfunctions to 
phase space are then, e.g., provided by their Wigner transforms. The choice 
of the projected eigenvectors $\cP_\nu\psi_j$ leads to consider expectation 
values of diagonal blocks $\cP_\nu \cB\cP_\nu$ of operators $\cB$ with symbols
$B\in\scl^q(1)$. On the symbol level the time evolution of these blocks
is covered by the Egorov theorem \ref{thm:Egorov}. Representing then the
blocks of the principal symbols by Stratonovich-Weyl symbols as described
in section \ref{sec:eigendyn}, according to Proposition \ref{prop:SWEgorov}
we are faced with the skew-product flows $Y^t_\nu$ on the product phase
spaces $\TRd\times\cO_\la$. Since the Stratonovich-Weyl symbols
$b_{0,\nu}$ defined in equation (\ref{eq:SWsymbdef}) that are associated 
with symbols $B\in\scl^q(1)$ are clearly integrable with respect to the 
measures $\dl\,\ud\eta$ on the (compact) manifolds $\Om_{\nu,E}\times\cO_\la$,
the (assumed) ergodicity of the flow $Y^t_\nu$ implies that
\begin{equation}
\label{eq:Ytergodb0}
\begin{split}
 \lim_{T\to\infty} \frac{1}{T} \int_0^T \bigl(b_{0,\nu} \circ Y^t_\nu \bigr)
 (x,\xi,\eta) \ \ud t 
 &= \frac{1}{\vol\cO_\la} \int_{\Om_{\nu,E}}\int_{\cO_\la} b_{0,\nu}
    (x',\xi',\eta')\ \ud\eta'\,\dl(x',\xi') \\
 &= \text{M}_{E,\nu,\la}(b_{0,\nu})
\end{split}
\end{equation}
holds for almost all initial conditions 
$(x,\xi,\eta)\in\Om_{\nu,E}\times\cO_\la$. In particular, one immediately
realises that the supposed ergodicity of $Y^t_\nu$ implies ergodicity for
the flow $\Phi^t_\nu$ on $\Om_{\nu,E}$ with respect to Liouville measure
$\dl$. As a consequence the condition (H6$_\nu$) is automatically
fulfilled.

For the subsequent formulation and proof of quantum ergodicity we choose to
follow in principle the approach of \cite{Zel96,ZelZwo96}. This means that
we investigate the variance of expectation values about their mean in the
semiclassical limit. In order to avoid the problem of explicitly estimating 
the norms of projected eigenvectors we here consider the normalised vectors 
$\phi_{j,\nu}$, defined in (\ref{eq:Defquasim}), which have been identified
as quasimodes for both the operators $\cH$ and $\cH\cP_\nu$. Moreover, we 
concentrate on vectors corresponding to projected eigenvectors with norms 
that do not vanish semiclassically, i.e., with $\|\cP_\nu \psi_j \|^2 \ge \de$ 
for some fixed $\de\in (0,1)$. This approach is similar to the one
introduced by Schubert \cite{Sch01} in the context of local quantum 
ergodicity, where an equidistribution was shown for quasimodes associated 
with ergodic components of phase space. In section \ref{sec:TraceAsymptotics} 
we estimated the relative number $N_{\nu,I}^\de/N_I$ of the associated
eigenvectors among all eigenvectors of $\cH$ in the semiclassical limit
from below, see (\ref{eq:Nde/N}). A non-trivial bound could only be obtained
for $\tilde\de<1$ corresponding to 
\begin{equation*}
 \de<\de_\nu:=\frac{k_\nu \vol\Om_{\nu,E}}{\sum_{\mu=1}^l 
 k_\mu\vol\Om_{\mu,E}}. 
\end{equation*}
Therefore, from now on we confine $\de$ to the interval $\de\in(0,\de_\nu)$,
and are thus in a position to state our main result.
\begin{theorem}
\label{thm:QE}
Let $\cH$ be a pseudodifferential operator with hermitian symbol 
$H\in\scl^0(m)$ whose principal part $H_0$ fulfills the conditions (H1) and 
(H2) of section \ref{sec:Egorov}. The eigenvalues $\la_1,\ldots,\la_l$ of
$H_0$ are required to have constant multiplicities and shall obey the 
conditions (H3$_\nu$)--(H5$_\nu$) of section \ref{sec:TraceAsymptotics} for 
all $\nu\in\{1,\ldots,l\}$. Moreover, they shall be separated according to the 
hyperbolicity condition (H0),
\begin{equation*}
 |\lambda_{\nu}(x,\xi) - \lambda_\mu(x,\xi) | \ge C m(x,\xi) \quad 
 \text{for} \quad \nu \neq \mu \quad \text{and} \quad |x|+|\xi| \ge c.
\end{equation*}
Assume now that the symbol $H\sim\sum_{j=0}^\infty \hbar^j H_j$ 
satisfies the growth condition
\begin{equation} 
\tag{\ref{eq:HamiltonianGrowth}}
  \| H_j\,^{(\alpha)}_{(\beta)}(x,\xi) \|_{n\times n} \le C_{\al,\be} \quad 
 \text{for all} \quad (x,\xi) \in \TRd \ \text{and} \ 
 |\alpha|+|\beta|+j \ge 2-\delta_{j0},
\end{equation}
and that the condition (Irr$_\nu$) of section \ref{sec:eigendyn} holds. 
If then the flow $Y^t_\nu$ defined in 
(\ref{eq:skewflowO}) is ergodic on $\Om_{\nu,E}\times\cO_\la$ with respect 
to the invariant measure $\dl\,\ud\eta$, in every sequence of normalised 
projected eigenvectors $\{ \phi_{j,\nu}\}_{j \in \nz}$, with 
$\| \cP_{\nu} \psi_j \|^2 \ge \de$, $\de \in (0,\de_\nu)$ fixed, one finds 
a subsequence $\{ \phi_{j_\alpha,\nu}\}_{\alpha \in \nz}$ of density one, 
i.e.,
\begin{equation*}
 \lim_{\hbar \to 0} \frac{\# \{\alpha; \ \| \cP_\nu \psi_{j_\alpha} 
 \|^2 \ge \delta \}}{\# \{j;\ \| \cP_\nu \psi_{j}\|^2 \ge \delta \}} =1,
\end{equation*}
such that for every operator $\cB$ with symbol $B \in \scl^0(1)$ and 
principal symbol $B_0$
\begin{equation} 
\label{eq:QEconv}
 \lim_{\hbar \to 0} \langle \phi_{j_\alpha,\nu}, \cB 
 \phi_{j_\alpha,\nu} \rangle = {\mathrm M}_{E,\nu,\lambda}(b_{0,\nu}),
\end{equation}
where $b_{0,\nu}$ denotes the Stratonovich-Weyl symbol associated with 
$P_{\nu,0} B_0 P_{\nu,0}$. Furthermore, the density-one subsequence 
$\{ \phi_{j_\alpha,\nu} \}_{\alpha \in \nz}$ can be chosen to be independent 
of the operator $\cB$.
\end{theorem}
\begin{proof}
We start with considering expectation values of the operator $\cB$ taken
in the quasimodes $\{ \phi_{j,\nu} \}$ and denote their variance about the 
mean $\mathrm{M}_{E,\nu,\lambda}(b_{0,\nu})$ of the corresponding 
Stratonovich-Weyl symbol $b_{0,\nu}$ defined in (\ref{eq:SWsymbdef}) as
\begin{equation*} 
 S_{2,\nu}^\delta(E,\hbar):= \frac{1}{N_{\nu,I}^\delta} 
 \sum_{\substack{E_j \in I(E,\hbar) \\ \| \cP_\nu \psi_j \|^2 \ge 
 \delta}} \bigl| \langle \phi_{j,\nu}, \cB \phi_{j,\nu} \rangle 
 - \mathrm{M}_{E,\nu,\lambda}(b_{0,\nu}) \bigr|^2.
\end{equation*}
Due to the definition (\ref{eq:Defquasim}) of the normalised vectors 
$\phi_{j,\nu}$, this variance can also be written as
\begin{equation*} 
\begin{split} 
 S_{2,\nu}^\delta(E,\hbar) 
 & = \frac{1}{N_{\nu,I}^\delta} \sum_{\substack{
     E_j \in I(E,\hbar) \\ \| \cP_\nu \psi_j \|^2 \ge \delta}}
     \bigl| \langle \phi_{j,\nu}, \bigl( \cB - \mathrm{M}_{E,\nu,\la}
     (b_{0,\nu}) \bigr) \phi_{j,\nu} \rangle \bigr|^2 \\
 &= \frac{1}{N_{\nu,I}^\delta} \sum_{\substack{ E_j \in I(E,\hbar) \\ 
    \| \cP_\nu \psi_j \|^2 \ge \delta}} 
    \| \cP_\nu \psi_j \|^{-2} \left| \langle \psi_j,\cP_\nu 
    \bigl(\cB - \mathrm{M}_{E,\nu,\la}(b_{0,\nu})\bigr) \cP_{\nu} 
    \psi_j \rangle \right|^2.
\end{split} 
\end{equation*}
Allowing for an error of $O(\hbar^\infty)$, in this expression the
expectation values can be replaced by those of the operator 
$\tilde\cP_\nu( \cB -\mathrm{M}_{E,\nu,\lambda}(b_{0,\nu}))\tilde\cP_\nu$ 
whose symbol is in the invariant subalgebra $\sinv^0(1)\subset\scl^0(1)$. 
Therefore, since all further requirements are also met, the Egorov theorem 
\ref{thm:Egorov} applies and yields that for finite times $t \in [0,T]$ the 
evolution $\cU^\ast(t)\tilde\cP_\nu (\cB -\mathrm{M}_{E,\nu,\lambda}
(b_{0,\nu}))\tilde\cP_\nu \cU(t)$ of this operator is again a 
pseudodifferential operator with symbol in the class $\scl^0(1)$. Taking 
into account that the $\psi_j$s are eigenvectors of $\cH$ with eigenvalues 
$E_j$, the above expression can be rewritten as
\begin{equation*}
 S_{2,\nu}^\delta(E,\hbar) = \frac{1}{N_{\nu,I}^\delta} 
 \sum_{\substack{E_j \in I(E,\hbar) \\ \| \cP_\nu \psi_j \|^2 \ge \delta}}
 \bigl| \langle \psi_j, \cB_{\nu,T} \psi_j \rangle \bigr|^2 \|\cP_\nu 
 \psi_j \|^{-2},
\end{equation*}
where we have defined the auxiliary operator 
\begin{equation} 
\label{eq:AuxOp}
 \cB_{\nu,T}:= \frac{1}{T} \int_0^T \cU^\ast(t) \tilde\cP_\nu 
 \bigl( \cB -\mathrm{M}_{E,\nu,\lambda}(b_{0,\nu}) \bigl) 
 \tilde\cP_\nu \cU(t)\ \ud t.
\end{equation}
Furthermore, by using using the Cauchy-Schwarz inequality
and the lower bound on the norms $\| \cP_\nu \psi_j \|^2 \ge \delta >0$
we obtain as an upper bound
\begin{equation*} 
 S_{2,\nu}^\delta(E,\hbar) \le \frac{1}{\delta} \frac{N_I}{N_{\nu,I}^\delta}
 \frac{1}{N_I}\sum_{E_j \in I(E,\hbar)} 
 \langle \psi_j, \cB_{\nu,T}^2 \psi_j \rangle.
\end{equation*}
According to equation (\ref{eq:Nde/N}) the factor $N_I/N_{\nu,I}^\delta$
can be estimated from above in the semiclassical limit. We hence now consider
the semiclassical limit of the expression 
\begin{equation*}
 \frac{1}{N_I} \sum_{E_j \in I(E,\hbar)} \langle \psi_j,
 \cB_{\nu,T}^2 \psi_j \rangle,
\end{equation*}
to which Proposition \ref{prop:Szego} can be applied. To this end one 
requires the principal symbol $B_{\nu,T,0}$ of the auxiliary operator 
$\cB_{\nu,T}$, which follows from Theorem \ref{thm:Egorov} as
\begin{equation*} 
 B_{\nu,T,0} = \frac{1}{T}\int_0^T  d_{\nu\nu}^\ast \bigl( (P_{\nu,0}
 B_0 P_{\nu,0}) \circ\Phi^t_\nu \bigr) d_{\nu\nu}\ \ud t - 
 \mathrm{M}_{E,\nu,\lambda}(b_{0,\nu})P_{\nu,0}.
\end{equation*}
Given this, the limit formula (\ref{eq:Szego}) and the estimate 
(\ref{eq:Nde/N}) yield
\begin{equation} 
\label{eq:S2Scl} 
\begin{split} 
 \lim_{\hbar \to 0} S_{2,\nu}^\delta(E,\hbar) 
 &\le \frac{1}{\de}\ \frac{\sum_{\mu=1}^l k_\mu \vol \Om_{\mu,E}}
   {(1-\tilde \delta) k_\nu \vol \Om_{\nu,E}} \ \frac{\vol\Om_{\nu,E} 
   \mtr\ell_{\nu,E}(B_{\nu,T,0}^2)}{\sum_{\mu=1}^l k_\mu 
   \vol\Om_{\mu,E}} \\
 &=\frac{1}{\de}\ \frac{1}{1- \tilde\de}\ \mathrm{M}_{E,\nu,\la}
   \bigl((\symb^{SW}[B_{\nu,T,0}])^2\bigr),
\end{split} 
\end{equation} 
when employing the tracial property $(v)$ of Proposition \ref{prop:SWsymbol}.

According to Proposition \ref{prop:SWEgorov} the Stratonovich-Weyl symbol 
of $B_{\nu,T,0}$ can now be easily calculated as
\begin{equation*}
 \symb^{SW}[B_{\nu,T,0}(x,\xi)](\eta) = \frac{1}{T} \int_0^T \bigl(
 b_{0,\nu} \circ Y^t_\nu\bigr)(x,\xi,\eta)\  \ud t - 
 \mathrm{M}_{E,\nu,\lambda}(b_{0,\nu}).
\end{equation*}
Since we assume the skew-product flow $Y^t_\nu$ to be ergodic with 
respect to $\dl\,\ud\eta$, the relation (\ref{eq:Ytergodb0}) implies 
that $\symb^{SW}[B_{\nu,T,0}(x,\xi)](\eta)$ vanishes in the limit $T\to\infty$
for almost all points $(x,\xi,\eta)\in\Om_{\nu,E}\times\cO_\la$. Now, on 
the right-hand side of (\ref{eq:S2Scl}) the square of 
$\symb^{SW}[B_{\nu,T,0}]$ enters integrated over $\Om_{\nu,E}\times\cO_\la$,
so that this expression vanishes as $T\to\infty$. We hence conclude that
\begin{equation*}
\lim_{\hbar \to 0} S_{2,\nu}^\delta(E,\hbar) = 0.
\end{equation*}
This, in turn, is equivalent to the existence of a subsequence 
$\{ \phi_{j_\alpha,\nu} \}_{\alpha\in\nz}\subset\{\phi_{j,\nu}\}_{j\in\nz}$ 
of density one, such that equation (\ref{eq:QEconv}) holds. Finally, by 
a diagonal construction as in \cite{Zel87,Col85} one can extract 
a subsequence of $\{ \phi_{j_\alpha,\nu}\}_{\alpha\in\nz}$ that is still
of density one in $\{ \phi_{j,\nu}\}_{j\in \nz}$, such that (\ref{eq:QEconv}) 
holds independently of the operator $\cB$.
\end{proof}
The version of quantum ergodicity asserted in Theorem \ref{thm:QE} means 
that in the semiclassical limit the lifts of almost all quasimodes 
$\phi_{j,\nu}$ to the phase space $\TRd\times\cO_\la$ equidistribute
in the sense that suitable Wigner functions (weakly) converge to an
invariant measure on $\Om_{\nu,E}\times\cO_\la$ that is proportional
to $\ud\ell\,\ud\eta$. In order to identify the proper Wigner transform
consider
\begin{equation*}
 \langle \phi_{j_\al,\nu},\cB\phi_{j_\al,\nu} \rangle = \frac{1}{(2\pi\hbar)^d}
 \iint_{\TRd} \mtr \Bigl( W[\phi_{j_\al,\nu}](x,\xi)P_\nu(x,\xi)B(x,\xi)
 P_\nu(x,\xi) \Bigr) \ \ud x\,\ud\xi +O(\hbar^\infty),
\end{equation*}
with the matrix valued Wigner transform
\begin{equation*}
 W[\psi](x,\xi) := \int_{\rz^d}\ue^{-\frac{\ui}{\hbar}\xi y} \overline{\psi}
 (x-\tfrac{y}{2}) \otimes \psi(x+\tfrac{y}{2})\ \ud y
\end{equation*}
defined for $\psi\in L^2(\rz^d)\otimes\kz^n$. We now exploit the 
Stratonovich-Weyl calculus to conclude that on the level of principal
symbols 
\begin{equation*}
\begin{split}
 \mtr \Bigl( W[\phi_{j_\al,\nu}]P_{\nu,0} B_0 P_{\nu,0} \Bigr)
 &= \mtr \Bigl( \bigl( V_\nu^\ast W[\phi_{j_\al,\nu}]V_\nu \bigr) \bigl(
    V^\ast_\nu B_0 V_\nu \bigr) \Bigr) \\
 &= \int_{\cO_\la} \symb^{SW}[V_\nu^\ast W[\phi_{j_\al,\nu}]V_\nu](\eta)
    \symb^{SW}[V_\nu^\ast B_0 V_\nu](\eta)\ \ud\eta.
\end{split}
\end{equation*}
The second factor in the integral has been defined as $b_{0,\nu}$ in
(\ref{eq:SWsymbdef}). In analogy to this we therefore introduce for 
$\psi\in L^2(\rz^d)\otimes\kz^n$ the scalar Wigner transform (see 
also \cite{BolGlaKep01})
\begin{equation*}
 w_\nu[\psi](x,\xi,\eta) := \symb^{SW}[V_\nu^\ast(x,\xi)W[\psi](x,\xi)
 V_\nu(x,\xi)](\eta),
\end{equation*}
that indeed provides a lift of $\psi$ to the phase space $\TRd\times\cO_\la$.
The statement of Theorem~\ref{thm:QE} can thus be rephrased in that under
the given conditions one obtains (in the sense of a weak convergence),
\begin{equation*}
 \lim_{\hbar\to 0}\frac{1}{(2\pi\hbar)^d}\ w_\nu[\phi_{j_\al,\nu}](x,\xi,\eta)
 \ \ud x\,\ud\xi\,\ud\eta =  \frac{1}{\vol\cO_\la}\,\ud\ell(x,\xi)\,\ud\eta
\end{equation*}
along the subsequence of density one. However, since in $\phi_{j,\nu}$ the 
normalisation of $\cP_\nu\psi_j$ is hidden, an equivalent equidistribution 
for the lifts of the projected eigenvectors is only shown up to a constant. 
In analogy to the discussion in \cite{Sch01} this means that in the sequence 
$\{\psi_j;\ E_j\in I(E,\hbar)\}$ there exists a subsequence $\{\psi_{j_\al}\}$
of density one such that as $\hbar\to 0$,
\begin{equation*}
\langle \psi_{j_\al},\cP_\nu\cB\cP_\nu\psi_{j_\al} \rangle =
\| \cP_\nu\psi_{j_\al} \|^2 \mathrm{M}_{E,\nu,\lambda}(b_{0,\nu}) + o(1),
\end{equation*}
with a corresponding statement for the scalar Wigner transforms 
$w_\nu[\cP_\nu\psi_{j_\al}]$. Notice that the factor 
$\| \cP_\nu\psi_{j_\al} \|^2$ is independent of the operator $\cB$ so that 
the subsequence can again be chosen independently of $\cB$. Therefore, a 
non-vanishing semiclassical limit only exists for those subsequences along 
which the norms $\| \cP_\nu\psi_{j_\al} \|$ do not tend to zero as 
$\hbar\to 0$. These subsequences are excluded in the formulation of 
Theorem~\ref{thm:QE} since $\de$ is fixed and positive.

The difficulties with estimating norms of the projected eigenvectors
$\cP_\nu\psi_j$ arise from the presence of several level surfaces
$\Om_{\nu,E}$ on which the lifts of eigenfunctions potentially condense 
in the semiclassical limit. The situation simplifies considerably, if
at the energy $E$ all of the $l$ level surfaces except one are empty. 
\begin{cor}
\label{cor:QEsingle}
If under the conditions stated in Theorem~\ref{thm:QE} only the level
surface $\Om_{\nu,E}\subset\TRd$ is non-empty, there exists a subsequence
$\{\psi_{j_\al}\}$ of density one in $\{\psi_j;\ E_j\in I(E,\hbar)\}$,
independent of the operator $\cB$, such that
\begin{equation*}
 \lim_{\hbar \to 0} \langle \psi_{j_\al}, \cP_\mu \cB \cP_\mu
 \psi_{j_\al} \rangle = \de_{\mu\nu}\,{\mathrm M}_{E,\nu,\la}(b_{0,\nu}).
\end{equation*}
\end{cor}
In this situation the norms $\| \cP_\mu\psi_{j_\al} \|$ converge to one
for $\mu=\nu$ and to zero otherwise as $\hbar\to 0$ along the subsequence.
The lifts of the eigenvectors therefore condense on the only available
level surface in $\TRd$, as one clearly would have expected.
\begin{rem} 
\label{rem:QEwithG}
As a condition for quantum ergodicity to hold we have assumed
the skew-product flow $Y^t_\nu$ on $\Om_{\nu,E}\times\cO_\la$ to be ergodic.
The reason for introducing this flow was to formulate a genuinely classical
criterion in terms of a dynamics on the symplectic phase space 
$\TRd\times\cO_\la$. The formulation will be somewhat simpler, if one 
refrains from insisting on a completely classical description and employs
the skew-product flow $\tilde Y^t_\nu$ defined on $\TRd\times G$, see 
(\ref{eq:skewflowG}), instead. Then the use of the Stratonovich-Weyl calculus
can be avoided. Such a formulation is based on a hybrid of the classical
Hamiltonian flow $\Phi^t_\nu$ on $\TRd$ and the dynamics represented
by the conjugation with the unitary matrices $D_\nu$, which appears to be 
quantum mechanical in nature. Both formulations, however, are equivalent in
the sense that, first, the Stratonovich-Weyl calculus relates the quantum 
dynamics in the eigenspace to a classical dynamics on the coadjoint orbit 
in a one-to-one manner. Second, in appendix \ref{app:skewprod} we show that
the skew-product $Y^t_\nu$ on $\Om_{\nu,E}\times\cO_\la$ is ergodic, if 
and only if the skew-product $\tilde Y^t_\nu$ is ergodic on 
$\Om_{\nu,E}\times G$. One can therefore formulate Theorem~\ref{thm:QE}
without recourse to the Stratonovich-Weyl calculus once the limit 
$\mathrm{M}_{E,\nu,\la}(b_{0,\nu})$ is expressed as 
\begin{equation*}
\text{M}_{E,\nu,\la} \bigl( b_{0,\nu} \bigr) = \frac{1}{k_\nu}
 \mtr \ell_{\nu,E}( P_{\nu,0} B_0 P_{\nu,0}),
\end{equation*}
see (\ref{eq:Mmeandef}). Up to equation (\ref{eq:S2Scl}) the proof of 
Theorem~\ref{thm:QE} proceeds in the same manner as shown. From this point 
on one can then basically follow the method of \cite{BolGla00}, and to this 
end represents the principal symbol $B_{\nu,T,0}$ of the auxiliary operator 
(\ref{eq:AuxOp}) in terms of the isometries $V_\nu$,
\begin{equation*} 
 V_\nu^\ast B_{\nu,T,0} V_\nu = \frac{1}{T} \int_0^T D_\nu^\ast 
 \bigl( (V_\nu^\ast B_0 V_\nu) \circ \Phi^t_\nu \bigr) D_\nu \ \ud t - 
 \frac{1}{k_\nu} \mtr \ell_{\nu,E}(V_\nu^\ast B_0 V_\nu).
\end{equation*}
We now suppose that the flow $\tilde Y^t_\nu$ is ergodic on 
$\Om_{\nu,E}\times G$ and choose the function $F(x,\xi,g):=\rho(g)^\ast
(V_\nu^\ast B_0 V_\nu)(x,\xi) \rho(g) \in L^1 (\Om_{\nu,E}\times G)\otimes
\mat_{k_\nu}(\kz)$ to exploit the ergodicity. This yields for almost
all initial values $(x,\xi,g)\in \Om_{\nu,E}\times G$ that 
\begin{equation*}
\begin{split}
 \lim_{T\to\infty} &\rho(g)^\ast V_\nu^\ast(x,\xi) B_{\nu,T,0}(x,\xi)
 V_\nu (x,\xi)\rho(g) \\  
 &= \int_{\Om_{\nu,E}}\int_G \rho(h)^\ast ( V_\nu^\ast B_0 V_\nu)(y,\zeta) 
    \rho(h) \ \ud h\,\dl(y,\zeta) - \frac{1}{k_\nu} \mtr \ell_{\nu,E} 
    (V_\nu^\ast B_0 V_\nu).
\end{split}
\end{equation*}
Furthermore, since the representation $(\rho,\kz^{k_\nu})$ is assumed to be 
irreducible and the integral in the above expression is invariant under 
conjugation with arbitrary elements of $U(k_\nu)$, Schur's lemma implies 
that this integral is a multiple of the identity in $\kz^{k_\nu}$, leading
to
\begin{equation*}
 \int_{\Om_{\nu,E}}\int_G \rho(h)^\ast ( V_\nu^\ast B_0 V_\nu)(y,\zeta) 
 \rho(h) \ \ud h\,\dl(y,\zeta)
 = \frac{1}{k_\nu} \mtr \ell_{\nu,E} (V_\nu^\ast B_0 V_\nu).
\end{equation*}
Due to the way the principal symbol $B_{\nu,T,0}$ enters on the right-hand 
side of (\ref{eq:S2Scl}), the conjugation with $V_\nu(x,\xi)\rho(g)$ as well 
as the restriction to almost all $(x,\xi,g)$ is inessential, so that again 
one concludes a vanishing of $S_{2,\nu}^\delta(E,\hbar)$ as $\hbar\to 0$. 
\end{rem}

%% file: app.poisson.tex
\section{Relations for Poisson brackets of matrix valued functions}
\label{app:MatrixPoisson}
In this appendix we collect some relations for Poisson brackets of 
matrix valued functions on the phase space $\TRd$ that are needed in
section \ref{sec:Egorov}. These relations are already stated in 
\cite{EmmWei96,GerMauMarPou97,Spo00} and can be verified by straightforward 
calculations.

Our convention for the Poisson bracket of smooth matrix valued functions
$A,B\in C^\infty (\TRd)\otimes \mat_n(\kz)$ is
\begin{equation*}
\{A,B\} := \partial_\xi A \, \partial_x B - \partial_x A \, \partial_\xi B.
\end{equation*}
The first general relation then reads
\begin{equation} 
\label{eq:ABC-relation}
 A \{B,C\} - \{A,B\}C = \{AB,C\}-\{A,BC\}.
\end{equation}
Furthermore, for the projection matrices $P=P P$ one finds
\begin{equation} 
\label{eq:PlPP}
 P \{ \lambda, P \} P = 0,
\end{equation}
where $\la$ is any smooth scalar function on $\TRd$.

For $B \in C^\infty(\TRd)\otimes\mat_n(\kz)$ commuting with $P$ one then 
derives
\begin{equation} 
 P \{\lambda,B\} P = \{\lambda,PBP\} - [PBP,[P,\{\lambda,P\}]],
\end{equation}
In particular, using (\ref{eq:ABC-relation}) for projection 
matrices one obtains
\begin{equation*}
 P \{ P,B \} - \{P,P\} B = \{P,B\} - \{P,PB\}
\end{equation*}
and
\begin{equation*}
 B\{P,P\} - \{B,P\}P = \{BP,P\}-\{B,P\}.
\end{equation*}
Using these relations together with the condition $[B,P]=0$ one gets 
\begin{equation*}
 P \bigl( \{B,P\}-\{P,B\} \bigr) P = [B,P\{P,P\}P].
\end{equation*}
Furthermore, for different projection matrices $P_\mu$ and $P_\nu$ with
$P_\mu P_\nu=0$ for $\nu \neq \mu$ the general relation 
(\ref{eq:ABC-relation}) implies
\begin{equation*}
 P_{\mu} \{P_\nu,P_\nu\}=-\{P_\mu,P_\nu\}(1-P_\nu)
\end{equation*}
and
\begin{equation*}
 \{P_\nu,P_\nu\} P_\mu = - (1-P_\nu) \{P_\nu,P_\mu\}.
\end{equation*}
In the case $[P_\nu,B]=0=[P_\mu,B]$ one finds
\begin{equation*} 
\begin{split} 
 P_\nu \{ P_\mu,B\} - \{P_\nu,P_\mu\} B & = - \{P_\nu,P_\mu B\}, \\
 B \{P_\mu,P_\nu\} - \{B,P_\mu\} P_\nu &= \{B P_\mu,P_\nu\}.
\end{split} 
\end{equation*}
These equations imply 
\begin{equation*}
 P_\nu \bigl( \{B,P_\mu\} - \{P_\mu,B\} \bigr) P_\nu
 = - [B,P_\nu \{ P_\mu, P_\mu \} P_\nu].
\end{equation*}
One can now apply the above relations to expressions of the type arising 
in section \ref{sec:Egorov}, i.e.,
\begin{equation*} 
\begin{split} 
 & P_\mu \left( \frac{\dpr}{\dpr t} B(t) + \frac{1}{2}
 \Bigl( \{B(t),\lambda_\nu P_\nu\} - \{\lambda_\nu P_\nu, B(t)\} \Bigr)
 + \ui [B(t),H_1] \right) P_\mu \\ & \quad = \frac{\dpr}{\dpr t}
 P_\mu B(t) P_\mu - \delta_{\nu \mu} \{ \lambda_\nu, P_\mu B(t) P_\mu \}
 \\ & \qquad  
 + \left[ \frac{\lambda_\nu}{2} (-1)^{\delta_{\nu \mu}} P_\mu \{ P_\nu, 
 P_\nu \} P_\mu - \delta_{\nu \mu} [P_\nu, \{\lambda_\nu, P_\nu\}] - \ui
 P_\mu H_1 P_\mu , P_\mu B(t) P_\mu \right].
\end{split} 
\end{equation*}
Therefore, the definition
\begin{equation} 
\label{eq:H1Tilde}
 \tilde H_1:= \ui (-1)^{\delta_{\nu \mu}} 
 \frac{\lambda_\nu}{2} P_\mu \{ P_\nu, P_\nu \} P_\mu
  - \ui \delta_{\nu \mu} [P_\nu,\{\lambda_\nu,P_\nu\}] + P_\mu H_1 P_\mu
\end{equation}
allows to conclude that
\begin{equation} 
\label{eq:ModBgl}
 \frac{\dpr}{\dpr t} P_\mu B P_\mu - \delta_{\nu \mu} \{ \lambda_\nu,
 P_\mu B P_\mu \} - \ui [\tilde H_1, P_\mu B P_\mu] = 0.
\end{equation}

%% file: app.skewprod.tex
\section{A relation between the ergodicity of two skew-pro\-duct flows}
\label{app:skewprod}
In section \ref{sec:eigendyn} we considered two types of skew-product 
dynamics built over the Hamiltonian flows $\Phi^t_\nu$ on $\TRd$. Both 
derive from the dynamics in the eigenvector bundles $E^\nu \to\TRd$ given 
by conjugating the diagonal blocks of principal symbols with the transport 
matrices $d_{\nu\nu}$ along integral curves of the Hamiltonian flows. After 
having fixed local orthonormal bases in the fibres, or isometries 
$V_\nu(x,\xi):\kz^{k_\nu}\to E^\nu (x,\xi)$, respectively, the transport
matrices $d_{\nu\nu}$ have been represented by unitary $k_\nu\times k_\nu$ 
matrices $D_\nu$, leading to the skew-product flows $\hat Y^t_\nu$ on 
$\TRd\times\U(k_\nu)$. We then noticed that the dynamics in the fibres
might not exhaust the whole group $\U(k_\nu)$, but only some subgroup
$G$, which is then represented in $\U(k_\nu)$. This led us to consider
the skew-product flows $\tilde Y^t_\nu$ on $\TRd\times G$, given as
$\tilde Y^t_\nu (x,\xi,g) = (\Phi^t_\nu (x,\xi),g_\nu(x,\xi,t)g)$, see
(\ref{eq:skewG}) and (\ref{eq:skewflowG}). Assuming that the representation 
$\rho$ of $G$ in $\U(k_\nu)$ is irreducible, we constructed a 
representation of the fibre dynamics on the coadjoint orbit $\cO_\la$ of
$G$ determined by $\rho$. We thus arrived at the skew-product flows 
$Y^t_\nu$ on the symplectic phase spaces $\TRd\times\cO_\la$, with
$Y^t_\nu (x,\xi,\eta) = (\Phi^t_\nu (x,\xi),\Ad^\ast_{g_\nu(x,\xi,t)}\eta)$,
see (\ref{eq:skewO}) and (\ref{eq:skewflowO}). In section \ref{sec:EquiDist}
we required either the flows $\tilde Y^t_\nu$ or $Y^t_\nu$, restricted
to the level surfaces $\Om_{\nu,E}\subset\TRd$ in the base manifold, to
be ergodic relative to the respective invariant measures $\dl\,\ud g$
or $\dl\,\ud\eta$. We now show:
\begin{prop} 
\label{lem:YTildeandY}
The flow $\tilde Y^t_\nu:\Om_{\nu,E} \times G \to \Om_{\nu,E} \times G$ 
is ergodic with respect to $\dl\,\ud g$, if and only if the associated 
flow $Y^t_\nu:\Om_{\nu,E} \times \cO_\nu \to \Om_{\nu,E} \times \cO_\nu$ is 
ergodic with respect to $\dl\,\ud\eta$.
\end{prop}
\begin{proof}
A convenient characterisation for the ergodicity of a flow $\Phi^t$ on a 
probability space $(\Sigma,\ud\text{m})$ with invariant measure $\ud\text{m}$ 
employs the 
flow-invariant subsets of $\Sigma$: The flow is ergodic with respect to 
$\ud\text{m}$, if and only if every measurable flow-invariant set has either 
measure zero or full measure. We now first consider the `if' direction 
asserted in the proposition and to this end assume that 
$Y^t_\nu$ on $\Om_{\nu,E}\times\cO_\la$ is ergodic with respect to 
$\dl\,\ud\eta$. Hence every measurable $Y^t_\nu$-invariant set 
$B\subset \Om_{\nu,E}\times\cO_\la$ has either measure zero or full measure. 
In order to relate these sets with subsets of $\Om_{\nu,E}\times G$ we
recall the composed map $G \stackrel{\pi}{\to} G/G_\la \stackrel{\kappa}{\to}
\cO_\la$ from section \ref{sec:eigendyn}, where $\pi$ denotes the canonical 
projection of $G$ onto $G/G_\la$ and $\kappa$ is the diffeomorphism that 
identifies $G/G_\la$ with $\cO_\la$. One then realises that the following 
diagram commutes:
\begin{equation} 
\label{eq:YTildeAndY}
\begin{diagram}
 \node{(x,\xi,g)} \arrow{e,t,T}{\tilde Y^t_\nu} 
 \arrow{s,l,T}{\id_{\TRd} \times \pi} 
 \node{\bigl(\Phi^t_\nu(x,\xi),g_\nu(x,\xi,t) g\bigr)} 
 \arrow{s,r,T}{\id_{\TRd} \times\pi} \\ 
 \node{(x,\xi,g G_\la)} \arrow{e,t,T}{\overline Y^t_\nu} 
 \arrow{s,l,T}{\id_{\TRd}\times\kappa} 
 \node{(\Phi^t_\nu(x,\xi),g_\nu(x,\xi,t) g G_\la)} 
 \arrow{s,r,T}{\id_{\TRd}\times\kappa} \\
 \node{(x,\xi,\eta)} \arrow{e,t,T}{Y^t_\nu} 
 \node{\bigl(\Phi^t_\nu(x,\xi),\Ad^\ast_{g_\nu(x,\xi,t)}(\eta)\bigr)}
\end{diagram},
\end{equation}
where $\overline Y^t_\nu$ is induced by $Y^t_\nu$ under 
$\id_{\TRd} \times \pi$.
According to this diagram a $\tilde Y^t_\nu$-invariant set 
$A \subset \Om_{\nu,E} \times G$ projects to a $Y^t_\nu$-invariant 
subset $(\id_{\TRd} \times \kappa\circ\pi)(A)$ of 
$\Om_{\nu,E} \times \cO_\la$. The assumed ergodicity of $Y^t_\nu$ then 
implies that the measure of $(\id_{\TRd} \times \kappa\circ\pi)(A)$ is zero 
or one. Now the normalised Haar measure $\ud g$ on $G$ projects under 
$\kappa\circ\pi$ to the volume measure $\ud\eta$ on the coadjoint orbit 
$\cO_\la$. This can be obtained from the Fubini theorem (cf. \cite{BroTom85}) 
which states for every $f \in L^1(\cO_\lambda)$ that
\begin{equation} 
\label{eq:Fubini} 
\begin{split} 
 \int_G (\pi^\ast \kappa^\ast f)(g)\ \ud g 
 &= \int_{G/G_\la} \left( \int_{G_\la} (\kappa^\ast f) \circ 
 \pi(g h)\ \ud h \right)\,\ud (g G_\la) \\
 & = \int_{G/G_\la} (\kappa^\ast f)(g G_\la)\ \ud (g G_\la).
\end{split} 
\end{equation}
Here $\ud h$ denotes the normalised Haar measure on $G_\la$ and 
$\ud (g G_\la)$ is the normalised left invariant volume form on $G/G_\la$ 
arising from the volume form on the coadjoint orbit under the pullback 
$\kappa^\ast$. Hence, the sets $A$ and $(\id_{\TRd} \times \kappa\circ\pi)(A)$
have identical measures and thus the measure of $A$ is either zero or one.
Therefore, the assumed ergodicity of $Y^t_\nu$ implies ergodicity of 
$\tilde Y_\nu^t$.

In order to prove the opposite direction one simply reverses the above
argument: Starting with $Y^t_\nu$-invariant subsets of $\Om_{\nu,E}\times
\cO_\la$, one lifts these to $\Om_{\nu,E}\times G$. Due to the commuting
diagram (\ref{eq:YTildeAndY}) these lifts are $\tilde Y^t_\nu$-invariant
and therefore, according to the assumed ergodicity of $\tilde Y^t_\nu$,
have measure zero or one. Again the Fubini theorem (\ref{eq:Fubini})
implies equal measures of the sets and their lifts. Hence $Y^t_\nu$
is ergodic.
\end{proof}